\documentclass[final,onefignum,onetabnum]{siamart0516} 
\usepackage{amsfonts,latexsym,verbatim,amscd,mathrsfs,color,array}
\usepackage{amsmath,amssymb,graphicx,color}
\usepackage{epstopdf}
\usepackage{diagbox}
\usepackage{subcaption}
\usepackage{multirow}

\def\theequation{\thesection.\@arabic\c@equation}
\makeatother
\renewcommand{\theequation}{\thesection.\arabic{equation}}

\newtheorem{remark}{Remark}[section]
\newtheorem{assumption}{Assumption}
\newcommand{\sech}{\text{sech}}

\newcommand{\dist}{\mathrm{dist}}
\newcommand{\crit}{\mathrm{crit}}
\newcommand{\BLSm}{\mathrm{BLS}_-}
\newcommand{\BLSp}{\mathrm{BLS}_+}

\newcommand{\ep}{\varepsilon}
\newcommand{\bl}{\mathrm{bl}}
\newcommand{\BL}{\mathrm{bl}}
\newcommand{\bls}{\mathrm{bls}}
\newcommand{\BLS}{\mathrm{BLS}}
\newcommand{\s}{\mathrm{s}}
\newcommand{\SPIKE}{\mathrm{s}}

\numberwithin{equation}{section}

\title{Existence and Stability of a Boundary Layer with an Interior Spike in the Singularly Perturbed Shadow Gierer-Meinhardt System}

 \author{Daniel Gomez \thanks{Center for Mathematical Biology \& Department of Mathematics, University of Pennsylvania,
 		Philadelphia, PA 19104, USA. (corresponding author {\tt d1gomez@sas.upenn.edu})} \and Juncheng Wei	\thanks{Department of Mathematics, University of British Columbia, Vancouver, BC V6T1Z2, Canada. {\tt
			jcwei@math.ubc.ca}}}

\begin{document}
	
\maketitle

\begin{abstract}
	The singularly perturbed Gierer-Meinhardt (GM) system in a bounded $d$-dimensional domain ($d\geq 2$) is known to exhibit boundary layer (BL) solutions for a non-zero activator flux. It was previously shown that such BL solutions can be destabilized by decreasing the activator flux below a stability threshold. Moreover, numerical simulations previously indicated that solutions consisting of a boundary layer and interior spike emerge after the destabilization of a BL solution. In this paper we use the method of matched asymptotic expansions to investigate the structure and stability of such ``boundary layer spike'' (BLS) solutions in the presence of an asymptotically small activator diffusivity $\ep^2\ll 1$. We find that two types of BLS solutions, one of which is unconditionally linearly stable and the other unstable, can be constructed provided that the activator flux is sufficiently small. In this way we determine that there is an asymptotically large range of activator flux values for which both the BL solution and one of the BLS solutions are linearly stable. Formal asymptotic calculations are further validated by numerically simulating the singularly perturbed GM system.
\end{abstract}


\section{Introduction}

An understanding of spatial patterns generated by reaction-diffusion equations modelling biological systems is a hallmark of mathematical biology. The aim of so-called \textit{toy models} is to incorporate only a few interactions so that the system remains analytically tractable and its results interpretable, while still retaining rich pattern forming behaviour reflecting that found in biological systems. The Gierer-Meinhardt (GM) system is one such model within which the pattern formation consequences of diffusion, activation, and inhibition can be investigated \cite{gierer_1972,meinhardt_2000}. Specifically, letting $u(x,t)$ and $\xi(x,t)$ denote the activator and inhibitor concentrations respectively the GM system takes the form of a two-component reaction-diffusion system. The GM system commonly takes the form
\begin{equation*}
	\frac{\partial u}{\partial t} = d\Delta u - u + \frac{u^2}{\xi},\qquad \tau\frac{\partial \xi}{\partial t} = D \Delta \xi - \xi + u^2,\qquad (x,t)\in\Omega\times(0,\infty),
\end{equation*}
where $d$ and $D$ denote the activator and inhibitor diffusivities respectively, and $\Omega\subset\mathbb{R}^N$ is a bounded domain on whose boundary $\partial\Omega$ additional conditions must be imposed. The GM system fits more broadly into the class of two-component reaction-diffusion systems exhibiting Turing instabilities \cite{turing_1952} such as the Gray-Scott, Schnakenberg, and Brusselator systems \cite{prigogine_1968,pearson_1993,schnakenberg_1979} (see also the textbook \cite{murray_2003}).

In the singularly perturbed limit for which $d\ll D$, the GM system is known to exhibit localized solutions in which the activator is concentrated in the vicinity of a discrete collection of points. Such solutions are often referred to as \textit{multi-spike} or \textit{multi-spot} solutions in $N=1$ or $N\geq 2$ dimensions respectively, and can also be found in other singularly perturbed reaction-diffusion systems \cite{nishiura_2002,ward_2018}. These localized solutions exhibit a separation of spatial and temporal scales which makes them particularly amenable to both formal and rigorous analysis \cite{takagi_1986,wei_2014_book}. Indeed, a substantial body of work has been devoted to studying the existence and stability of localized solutions to the singularly perturbed GM system and its various extensions \cite{doelman_2001_gm,iron_2001,maini_2007,gomez_2019}.

Studies of pattern formation in reaction-diffusion systems typically assume homogeneous Neumann, or no-flux, boundary conditions. The choice of no-flux boundary conditions is based in part on an underlying assumption that the system is closed or isolated from its environment. In addition such homogeneous boundary conditions provide a technical advantage as little or no additional assumptions are needed to guarantee that the system admits a spatially homogeneous steady state. This latter point is particularly important as it simplifies the analysis of Turing instabilities. However, it is increasingly apparent that different boundary conditions can have a substantial effect on pattern formation (see for example \cite{dillon_1994}). Inhomogeneous boundary conditions in particular arise naturally in heterogeneous problems \cite{krause_2020_isolating} as well as bulk-surface coupled systems \cite{levine_2005,madzvamuse_2015,ratz_2014,gomez_2021_rs,gomez_2019}.

In the context of localized solutions there is a small but growing body of literature considering boundary conditions deviating from standard homogeneous Neumann boundary conditions. Specifically, Maini et.\@ al.\@ considered in \cite{maini_2007} the stability of spikes in the shadow GM system under homogeneous Robin boundary conditions for both the activator and inhibitor (see also \cite{berestycki_2003} for an earlier analysis of the underlying half-space core problem). In addition, Tzou and Ward considered the effects of inhomogeneous inhibitor boundary conditions on the existence and stability of localized solutions to the singularly perturbed Brusselator model \cite{tzou_2018_brusselator}. Two additional studies which most closely inform our present paper are \cite{gomez_2021,gomez_2022} in which the authors considered inhomogeneous boundary conditions for the activator in the singularly perturbed GM system. Importantly, the asymptotically small diffusivity of the activator results in the formation of a boundary layer whose existence and linear stability was investigated in \cite{gomez_2022}. In particular it was found that when  $\Omega\subset\mathbb{R}^N$ with $N\geq 2$ the boundary layer is unstable when the boundary flux is sufficiently small. Numerical simulations further revealed the emergence of an interior spike after the destabilization of a boundary layer (see Figure 9 in \cite{gomez_2022}). This numerical observation serves as the primary motivation for this paper, in which we use the method of matched asymptotic expansions to construct and study the linear stability of these interior and near-boundary spike solutions.

\begin{figure}[t!]
	\begin{subfigure}{0.33\textwidth}
		\centering
		\includegraphics[width=\linewidth]{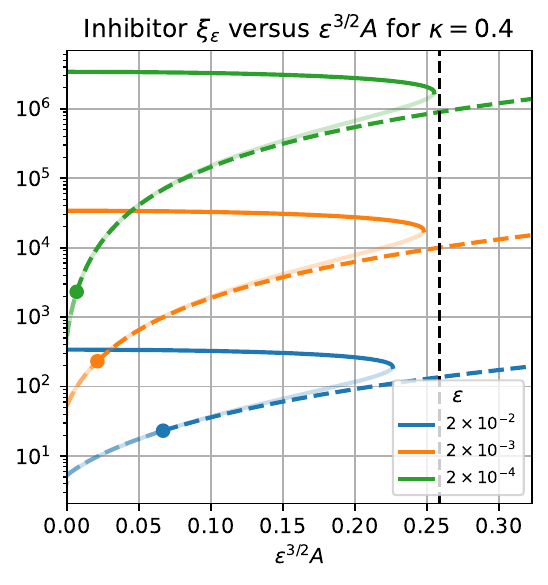}
	\end{subfigure}%
	\begin{subfigure}{0.33\textwidth}
		\centering
		\includegraphics[width=\linewidth]{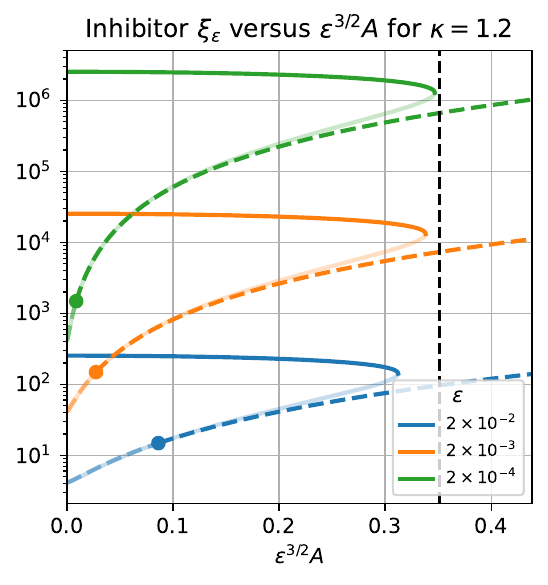}
	\end{subfigure}
	\begin{subfigure}{0.33\textwidth}
		\centering
		\includegraphics[width=\linewidth]{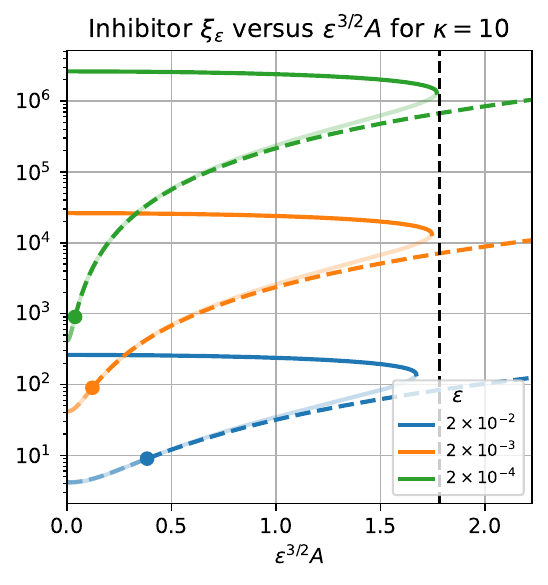}
	\end{subfigure}
	\caption{Plots of the inhibitor $\xi_\ep$ versus the rescaled activator flux $\ep^{3/2}A$ for (left) $\kappa=0.4$, (middle) $\kappa=1.2$, and (right) $\kappa=10$. Solid curves correspond to solutions consisting of a boundary layer with an interior spike, with the upper darkly coloured branch indicating the stable small-shift solution and the lower lightly coloured branch indicating the unstable large-shift solution. The dashed curves correspond to solutions consisting of only a boundary layer with the solid dot demarcating the region where is linearly stable (darkly coloured) and unstable (lightly coloured).}\label{fig:bifurc}
\end{figure}

Taking the inhibitor diffusivity $D\rightarrow\infty$ and appropriately rescaling variables we obtain the shadow GM system
\begin{subequations}\label{eq:gm}
	\begin{align}
		& \partial_t u = \ep^2 \Delta u - u + \frac{u^2}{\xi}, & x\in\Omega,\quad t>0, \label{eq:gm-u}\\
		&  \tau\xi_t = -\xi + \frac{1}{|\Omega|}\int_\Omega u^2dx, & t>0, \label{eq:gm-xi}\\
		& \ep\partial_\nu u + \kappa u = A,& x\in\partial\Omega,\quad t>0,  \label{eq:gm-bcs}
	\end{align}
\end{subequations}
where $0<\varepsilon\ll 1$ is an asymptotically small parameter, $\tau>0$, and $A>0$ is a scalar controlling the boundary flux. In this paper we will be interested in the existence and stability of two types of localized solutions. The first, which was previously considered in \cite{gomez_2022}, consists of a boundary layer concentrating along $\partial\Omega$ and we will refer to it as a boundary-layer (BL) solution. The second consists of a boundary layer and an interior spike and will be referred to as a boundary-layer-spike (BLS) solution which emerges in two types denoted by $\BLSm$ and $\BLSp$.  The primary contribution of this paper is the asymptotic analysis of the existence and linear stability of BLS solutions and is summarized in the following result.
\begin{principal_result*}
	Let $\varepsilon\ll 1$, $\tau\geq 0$, $\kappa\geq 0$, and $A > 0$. Let $w_c(y)$ be the one-dimensional homoclinic solution satisfying \eqref{eq:homoclinic-equation}. Additionally, let
	\begin{equation}
		\overline{W}_\kappa(y) := \begin{cases} W_\kappa(y) \quad \text{for } y\in\mathbb{R}^N_+:=\{(y_1,...,y_N)\in\mathbb{R}^N\,|\,y_N>0\},& \text{if }\kappa\leq \kappa_\star, \\ W(y)\quad\text{for }y\in\mathbb{R}^N, & \text{if }\kappa > \kappa_\star,\end{cases}
	\end{equation}
	and 
	\begin{equation}
		C_{N,\kappa} := \begin{cases}
			\int_{\mathbb{R}^N_+} W_\kappa(y)^2dy, & \kappa\leq \kappa_\star, \\
			\int_{\mathbb{R}^N} W(y)^2 dy, & \kappa>\kappa_\star,
		\end{cases}
	\end{equation}
	where $W_\kappa$ and $W$ are the unique least-energy solutions to \eqref{eq:half-space-core-problem} and \eqref{eq:full-space-core} respectively, and where $\kappa_\star>1$ is the unique threshold predicted by Theorem 1.1 of \cite{berestycki_2003} ($\lambda_*$ in their notation). Then, there exists a threshold $A=A_{\crit,\bls}^\ep>0$ with the limiting behaviour
	\begin{equation}\label{eq:princ-result-A-crit-limiting} 
		A_{\crit,\bls}^\ep \sim \frac{(1+\kappa)|\Omega|}{\sqrt{2|\partial\Omega|C_{N,\kappa}}}\ep^{-\frac{N+1}{2}},
	\end{equation}
	such that for all $0<A<A_{\crit,\bls}^\ep$ the singularly perturbed shadow GM system \eqref{eq:gm} admits two equilibrium solutions $(u,\xi)=(u_\ep^\pm,\xi_\ep^\pm)$ in which $u_\ep^\pm(x)$ consists of a boundary layer and an interior spike, and which are henceforth referred to as $\mathrm{BLS}_\pm$ solutions. Specifically
	\begin{equation}
		u_\ep^\pm(x)\sim \xi_\ep^\pm \left(w_c\left(\tfrac{\dist(x,\partial\Omega)}{\ep}+y_{\ep}^\pm \right) + \overline{W}_\kappa\left(\tfrac{x-x_0}{\ep}\right)\right),\qquad \xi_\ep^\pm \sim \tfrac{|\Omega|}{\ep |\partial\Omega|\eta(y_{\ep}^\pm) + \ep^{N} C_{N,\kappa}},
	\end{equation}
	where $y_{\ep}^\pm = -\log z_\pm$ and where $0<z_-<z_+$ are the unique positive solutions to the cubic
	\begin{equation}\label{eq:princ-result-cubic}
		q_\ep\biggl( 6z^2(z+3) + \ep^{N-1}\frac{C_{N,\kappa}}{|\partial\Omega|}(1+z)^3 \biggr) - 6z\bigl(1+\kappa - (1-\kappa)z\bigr) = 0,
	\end{equation}
	where $q_\ep := \ep A \frac{|\partial\Omega|}{|\Omega|}$. Moreover, if $\tau$ is sufficiently small then the $\BLSm$ solution is  linearly stable, whereas the $\BLSp$ solution is always linearly unstable.
\end{principal_result*}

In Figure \ref{fig:bifurc} we summarize the bifurcation structure of the BL and BLS solutions in $N$$=$$2$-dimensions by plotting the inhibitor $\xi$ versus $\ep^{3/2}A$. The solid curves correspond to the BLS solutions with the dark upper  (resp. light lower) component of each curve corresponding to the $\BLSm$ (resp. $\BLSp$) solution. On the other hand, the dashed curves correspond to the BL solution with the dark (resp. light) component indicating the regions where it is stable (resp. unstable). The solid dot in each plot indicates the point at which the BL solution changes stability and corresponds to a value that is $A=O(\ep^{-1})$ (see Section \ref{sec:bl_sol} below). Moreover, the dashed vertical line indicates the limiting behaviour of the existence threshold found in \eqref{eq:princ-result-A-crit-limiting}. Together with the results in \cite{gomez_2022} we draw the conclusions that if $A>0$ is sufficiently small then only the $\BLSm$ solution is linearly stable, whereas if $A>0$ is sufficiently large then only the BL solution exists and is linearly stable. Importantly, we also observe that there is a large range of $A$ values over which both the $\BLSm$ and BL solutions exist and are linearly stable.

The remainder of the paper is organized as follows. In Section \ref{sec:bl_sol} we summarize the existence and stability results found in \cite{gomez_2022} for the BL solution. In Section \ref{sec:bls_existence} we use the method of matched asymptotic expansions to calculate existence thresholds and construct equilibrium BLS solutions, while in Section \ref{sec:bls_stability} we consider their linear stability. We include in Section \ref{sec:numerical} a collection of numerical simulations validating our formal asymptotics while also suggesting that the destabilization of the BL solution leads to the emergence of the $\BLSm$ solution and vice versa. Throughout our calculations, a certain half-space core problem previously considered in \cite{berestycki_2003} and arising also in \cite{maini_2007} is prominently featured. In Appendix \ref{app:half_space_core} we numerically calculate solutions to this half-space core problem while in Appendix \ref{app:half_space_nlep} we consider its associated non-local eigenvalue problem.

\section{Boundary Layer Solutions and their Linear Stability}\label{sec:bl_sol}

In this section we summarize the partial results for the existence and linear stability of boundary layer solutions to \eqref{eq:gm} established in \cite{gomez_2022}. Let $w_c(y)$ be the unique homoclinic solution satisfying
\begin{equation}\label{eq:homoclinic-equation}
	\begin{cases}
		w_c'' - w_c + w_c^2 = 0,& -\infty<y<\infty, \\ 
		w_c'(0)=0\quad\text{and}\quad w_c(y)\rightarrow 0\quad \text{as}\quad y\rightarrow\pm\infty,\\
	\end{cases}
\end{equation}
Note that the solution is explicitly given by $w_c(y) = \tfrac{3}{2}\sech^2(y/2)$.  Using the method of matched asymptotic expansion, it can be shown that a boundary-layer solution to \eqref{eq:gm} is given by
\begin{equation*}
	u \sim \xi_{\ep,\BL} w_c\left(\varepsilon^{-1}\text{dist}(x,\partial\Omega) + y_{\ep,\BL}\right),\qquad \xi \sim \xi_{\ep,\BL} := \frac{1}{\ep}\frac{|\Omega|}{|\partial\Omega|}\frac{1}{\eta(y_{\ep,\BL})},
\end{equation*}
where
\begin{equation}\label{eq:eta-def}
	\eta(y_{\ep,\BL}):= \int_{y_{\ep,\BL}}^\infty w_c(y)^2dy,
\end{equation}
and the \textit{shift parameter} $y_{\ep,\BL}\in\mathbb{R}$ is chosen to satisfy the inhomogeneous boundary conditions
\begin{equation*}
	-w_c'(y_{\ep,\BL}) + \kappa w_c(y_{\ep,\BL}) = \ep A\frac{|\partial\Omega|}{|\Omega|}\eta(y_{\ep,\BL}).
\end{equation*}
whose solution is explicitly given by
\begin{equation*}
	y_{\ep,\BL} = \log\left(\frac{1-\kappa + 3q_\ep+\sqrt{(1-\kappa+3q_\ep)^2+4(1+\kappa)q_\ep}}{2(1+\kappa)}\right),
\end{equation*}
where $q_\ep = \ep A\tfrac{|\partial\Omega|}{|\Omega|}$. In Theorem 3.1 of \cite{gomez_2022} the authors rigorously established the existence and linear stability of the boundary layer solution for $A>A_{\text{crit},\bl}^\ep(\kappa)$ where
\begin{eqnarray}
	A_{\text{crit},\bl}^\ep(\kappa) := \frac{|\Omega|}{\ep|\partial\Omega|} \left(\frac{3-\kappa+\sqrt{\kappa^2+3}}{3+\kappa-\sqrt{\kappa^2+3}}\right)\left(\frac{2\kappa+\sqrt{\kappa^2+3}}{6-\kappa+\sqrt{\kappa^2+3}}\right).
\end{eqnarray}
Furthermore, numerical simulations suggest that the boundary layer solution is unstable for $A<A_{\text{crit},\bl}^\ep(\kappa)$ with the resulting instabilities leading to the formation of an interior spike (see Section 3.3 and Figure 9 of \cite{gomez_2022}). In the remainder of this paper we will use the method of matched asymptotic expansions to construct this interior spike solution and determine its linear stability.

\section{Asymptotic Construction of Boundary-Layer Solutions with an Interior or Near-Boundary Spike}\label{sec:bls_existence}

We seek an equilibrium solution to \eqref{eq:gm} consisting of a boundary layer and spike concentrated at an interior point. Specifically we decompose the solution as
\begin{subequations}
	\begin{equation}\label{eq:u_ep}
		u_\ep(x) = \xi_\ep \left( u_{\ep,\BL}\left(x\right) + u_{\ep,\SPIKE}\left(x\right)  \right),
	\end{equation}
	where $u_{\ep,\BL}(x)$ corresponds to a boundary-layer satisfying
	\begin{equation}\label{eq:bl-eq}
		\begin{cases}
			\ep^2\Delta u_{\ep,\BL} - u_{\ep,\BL} + u_{\ep,\BL}^2 = 0,& x\in\Omega,\\
			\ep\partial_\nu u_{\ep,\BL} + \kappa u_{\ep,\BL} = A/\xi_\ep, & x\in\partial\Omega,
		\end{cases}
	\end{equation}
	and $u_{\ep,\SPIKE}(x)$ corresponds to an interior spike satisfying
	\begin{equation}\label{eq:s-eq}
		\begin{cases}
			\ep^2\Delta u_{\ep,\SPIKE} -(1 -2 u_{\ep,\BL} )u_{\ep,\SPIKE} + u_{\ep,\SPIKE}^2 = 0,& x\in\Omega,\\
			\ep\partial_\nu u_{\ep,\SPIKE} + \kappa u_{\ep,\SPIKE} = 0, & x\in\partial\Omega.
		\end{cases}
	\end{equation}
\end{subequations}
Proceeding as in \cite{gomez_2022} we readily determine that the boundary-layer is given by
\begin{equation*}
	u_{\ep,\BL}(x) \sim w_0(x) := w_c\left(\frac{\dist(x,\partial\Omega)}{\ep} + y_{\ep,\bls} \right),
\end{equation*}
where $w_c(y)$ is the one-dimensional homoclinic solution satisfying \eqref{eq:homoclinic-equation}, and where the shift parameter $y_{\ep,\bls}$ will be determined by enforcing the inhomogeneous boundary condition.

In contrast to $u_{\ep,\BL}$, the interior spike solution $u_{\ep,\SPIKE}$ can be drastically different depending on the value of $\kappa\geq 0$. To understand why, it is instructive to first consider the problem
\begin{equation}\label{eq:homogeneous-bc-eq}
	\begin{cases}
		\ep^2\Delta U_{\ep,\kappa} - U_{\ep,\kappa} + U_{\ep,\kappa}^2 = 0, & x\in\Omega,\\
		\ep\partial_\nu U_{\ep,\kappa} + \kappa U_{\ep,\kappa} = 0, & x\in\partial\Omega,
	\end{cases}	
\end{equation}
for which we seek a spike solution concentrating at $x_\ep=\text{argmax}_{x\in\Omega}U_{\ep,\kappa}(x)$. In Theorems 1.1--1.3 of \cite{berestycki_2003} it was rigorously found that there exists a critical threshold $\kappa_\star>1$ such that as $\ep\rightarrow 0^+$:
\begin{enumerate}
	\item[(i)] If $\kappa \leq \kappa_\star$ then $\dist(x_\ep,\partial\Omega)\rightarrow \ep d_0$ for some $d_0>0$, $x_\ep\rightarrow x_0\in\partial\Omega$, and $U_{\ep,\kappa}(x_0 + \ep y)\rightarrow W_\kappa(y)$ in $C^1$ locally, where $W_\kappa(y)$ is the least-energy solution to the \textit{half-space core problem}
	\begin{subequations}
		\begin{equation}\label{eq:half-space-core-problem}
			\begin{cases}
				\Delta W_\kappa - W_\kappa + W_\kappa^2 = 0,\quad W_\kappa>0 & y\in\mathbb{R}^N_+ := \{(y_1,...,y_N)\in\mathbb{R}^N\,|\, y_N>0\},\\
				\partial_\nu W_\kappa + \kappa W_\kappa = 0, & y\in\partial\mathbb{R}^N_+.
			\end{cases}
		\end{equation}
		\item[(ii)] If $\kappa > \kappa_\star$ then $x_\ep \rightarrow  x_0 = \text{argmax}_{x\in\Omega}\dist(x,\partial\Omega)$ and $U_{\ep,\kappa}(x_\ep + \ep y)\rightarrow W(y)$ in $C^1$ locally, where $W(y)$ is the least-energy solution to the \textit{full-space core problem}
		\begin{equation}\label{eq:full-space-core}
			\begin{cases}
				\Delta W - W + W^2 = 0,\quad W>0 & y\in\mathbb{R}^N,\\
				W(0)=\max_{y\in\mathbb{R}^N} W(y),\quad\text{and}\quad W(y)\rightarrow 0\quad\text{as}\quad |y|\rightarrow\infty.
			\end{cases}
		\end{equation}
	\end{subequations}
\end{enumerate}
In each of the above cases, the least-energy solution refers to that which minimizes the energy
\begin{equation*}
	I_\kappa[u] = \int_{\mathbb{R}^N_+}\biggl(\frac{1}{2}|\nabla u|^2  + \frac{1}{2}u^2\biggr) - \frac{1}{p+1}\int_{\mathbb{R}^N_+}u^{p+1} + \frac{\kappa}{2}\int_{\mathbb{R}^N_+} u^2,
\end{equation*}
in case (i), and 
\begin{equation*}
	I[u] = \int_{\mathbb{R}^N_+}\biggl(\frac{1}{2}|\nabla u|^2  + \frac{1}{2}u^2\biggr) - \frac{1}{p+1}\int_{\mathbb{R}^N_+}u^{p+1},
\end{equation*}
in case (ii). We refer the reader to Appendix \ref{app:half_space_core}  for additional discussion on the numerical calculation of solutions to \eqref{eq:half-space-core-problem} and the threshold $\kappa_\star > 1$.

It is evident from the above discussion that the spike solution $u_{\ep,\SPIKE}$ may qualitatively change depending on whether $\kappa\leq \kappa_\star$ or $\kappa>\kappa_\star$, concentrating at a point that is an $O(\ep)$ or $O(1)$ distance from the boundary $\partial\Omega$ in each case respectively. In order to draw such a conclusion we compare \eqref{eq:s-eq} and \eqref{eq:homogeneous-bc-eq}, in light of which we make the following assumption on the shift parameter.

\begin{assumption}\label{ass:shift}
	There exists a positive constant $C=O(1)$ such that if $\kappa\leq\kappa_\star$ then the shift-parameter $y_{\ep,\bls}\gg C$ whereas if $\kappa>\kappa_\star$ then $y_{\ep,\bls}>-C$.
\end{assumption}

These assumptions simplify the subsequent asymptotic analysis by controlling the contribution of the boundary-layer $u_{\ep,\BL}(x)$ near the spike location $x_\ep = \text{argmax}_{x\in\Omega}u_{\ep,\s}(x)$.  Specifically, regardless of whether $\kappa\leq\kappa_\star$ or $\kappa>\kappa_\star$, under Assumption \ref{ass:shift} we will always have that $w_0(x_\ep+\ep y)\ll 1$  for all $y=O(1)$. Proceeding with the method of matched asymptotic expansions and noting that $(1-2u_{\ep,\BL}(x))\approx 1$ for $x$ near $x_\ep$, we then deduce that
\begin{equation}
	u_{\ep,\SPIKE}(x) \sim \overline{W}_\kappa\left(\frac{x-x_0}{\ep}\right) := \begin{cases}
		W_\kappa(\ep^{-1}(x-x_0)), & \kappa\leq\kappa_\star, \\
		W(\ep^{-1}(x-x_0)), & \kappa>\kappa_\star.
	\end{cases}
\end{equation}
where $W_\kappa(y)$ and $W(y)$ solve \eqref{eq:half-space-core-problem} and \eqref{eq:full-space-core} respectively. Defining $\eta(y_{\ep,\bls})$ by \eqref{eq:eta-def} and
\begin{equation}
	C_{N,\kappa} := \begin{cases}
		\int_{\mathbb{R}^N_+} W_\kappa(y)^2dy, & \kappa\leq \kappa_\star, \\
		\int_{\mathbb{R}^N} W(y)^2 dy, & \kappa>\kappa_\star,
	\end{cases}
\end{equation}
we thus obtain the following leading order approximation for the inhibitor
\begin{equation}\label{eq:xi_eps_def}
	\xi_\ep\sim \frac{|\Omega|}{\ep |\partial\Omega|\eta(y_{\ep,\bls}) + \ep^{N} C_{N,\kappa}}.
\end{equation}

\begin{figure}[t!]
	\begin{subfigure}{0.32\textwidth}
		\centering
		\includegraphics[width=\linewidth]{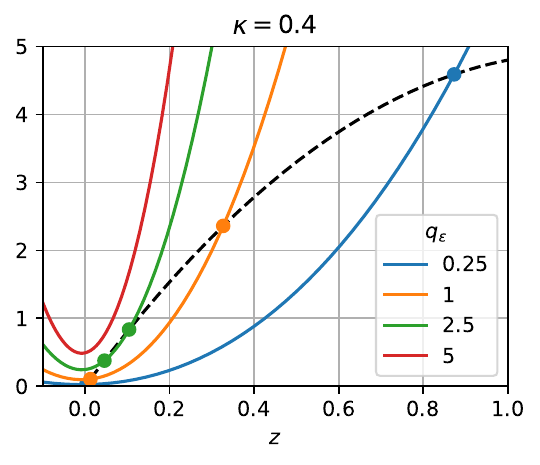}
	\end{subfigure}%
	\begin{subfigure}{0.32\textwidth}
		\centering
		\includegraphics[width=\linewidth]{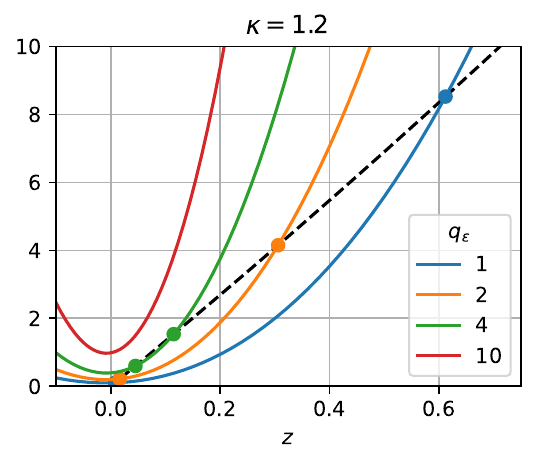}
	\end{subfigure}
	\begin{subfigure}{0.32\textwidth}
		\centering
		\includegraphics[width=\linewidth]{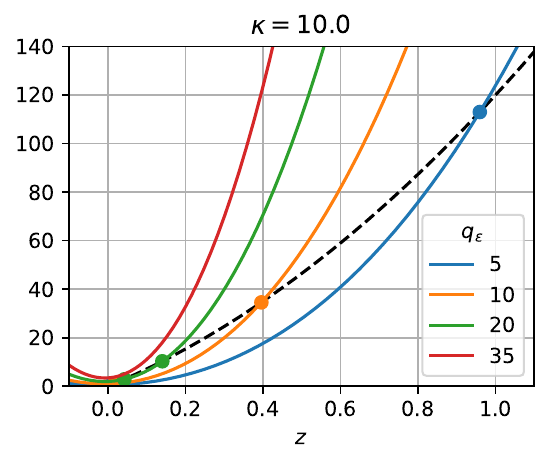}
	\end{subfigure}
	\caption{Plots of the left-hand-side (dashed) and right-hand-sides (solid) of the cubic equation \eqref{eq:cubic-original} for (left) $\kappa=0.4$, (middle) $\kappa=1.2$, and (right) $\kappa=10$. The plots illustrate the existence of a threshold for $q_\ep$ below which the cubic admits exactly two positive real roots, and beyond which it has none. In each plot $\ep=0.02$ and $N=2$.}\label{fig:example-zeros}
\end{figure}

The only remaining unknown in the preceding asymptotic construction is the shift parameter $y_{\ep,\bls}$ which is determined by enforcing the boundary condition in \eqref{eq:bl-eq}. Changing to boundary-fitted coordinates and retaining only the leading-order terms we find that $y_{\ep,\bls}$ solves
\begin{equation}\label{eq:shift-eq-bc}
	-w_c'(y_{\ep,\bls}) + \kappa w_c(y_{\ep,\bls}) = \frac{A}{|\Omega|}\left( \ep |\partial\Omega|\eta(y_{\ep,\bls}) + \ep^{N} C_{N,\kappa}\right).
\end{equation}
This nonlinear equation is readily rewritten as a cubic in the positive unknown $z=\exp(-y_{\ep,\bls})$ by noting that
\begin{equation}\label{eq:z-conversion}
	w_c(y_{\ep,\bls}) =  \frac{6z}{(1+z)^{2}},\quad w_c'(y_{\ep,\bls}) = -\frac{6z(1-z)}{(1+z)^{3}},\quad  \eta(y_{\ep,\bls}) = \frac{6z^2(3+z)}{(1+z)^{3}},
\end{equation}
with which \eqref{eq:shift-eq-bc} becomes
\begin{equation}\label{eq:cubic-original}
	6z\bigl(1+\kappa - (1-\kappa)z\bigr) = q_\ep\biggl( 6z^2(z+3) + \ep^{N-1}\frac{C_{N,\kappa}}{|\partial\Omega|}(1+z)^3 \biggr),\qquad q_\ep := \ep A\frac{|\partial\Omega|}{|\Omega|}.
\end{equation}
It is easy to see that there is an upper threshold for $q_\ep$ below which \eqref{eq:cubic-original} always has two positive solutions $0<z_-<z_+$, and above which it has no positive solutions. Since $\ep\ll 1$, we see that $z_-\ll 1$ whereas $z_+$ is bounded above by $\tfrac{1+\kappa}{1-\kappa}$ when $\kappa<1$ but may become arbitrarily large for $\kappa\geq 1$. Figure \ref{fig:example-zeros} illustrates these observations in which the dashed black curve indicates the left-hand-side of \eqref{eq:cubic-original} whereas the coloured curves correspond to the right-hand-side for different values of $q_\ep$ shown in the legend.

\subsection{Leading Order Behaviour of the Shift Parameter}\label{subsec:leading-order-shift}

In this subsection we determine a leading order expression for the shift parameter $y_{\ep,\bls}$ solving \eqref{eq:shift-eq-bc}. Let
\begin{equation}
	A = \ep^{\gamma-1}A_0,\qquad q_\ep = \ep^{\gamma} q_0,\qquad q_0 = A_0\frac{|\partial\Omega|}{|\Omega|},
\end{equation}
so that the cubic equation \eqref{eq:cubic-original} becomes
\begin{equation}\label{eq:cubic-full}
	\begin{split}
		\bigl(\tfrac{q_0 C_{N,\kappa}}{|\partial\Omega|}\ep^{\gamma+N-1} + 6q_0\ep^\gamma\bigr) z^3 + & \bigl( 3\tfrac{q_0 C_{N,\kappa}}{|\partial\Omega|}\ep^{\gamma+N-1} + 18q_0\ep^\gamma + 6(1-\kappa)  \bigr)z^2  \\
		+	& \bigl( 3\tfrac{q_0 C_{N,\kappa}}{|\partial\Omega|}\ep^{\gamma+N-1} - 6(1+\kappa) \bigr)z + \tfrac{q_0 C_{N,\kappa}}{|\partial\Omega|}\ep^{\gamma+N-1} = 0.
	\end{split}
\end{equation}
We seek strictly positive solutions to \eqref{eq:cubic-full} in three distinct cases: $\gamma>0$, $\gamma=0$, and $\gamma<0$. In each case, we consider only those solutions for which Assumption \ref{ass:shift} is satisfied. 

\vspace{2ex}\noindent\textbf{Case I:} Suppose that $\gamma > 0$. Neglecting higher order terms in \eqref{eq:cubic-full} we obtain
\begin{equation*}
	\underbrace{6q_0\ep^\gamma z^3}_{\mathrm{(I)}} + \underbrace{\bigl( 18q_0\ep^\gamma + 6(1-\kappa)  \bigr)z^2}_{\mathrm{(II)}} - \underbrace{6(1+\kappa) z}_{\mathrm{(III)}} + \underbrace{\tfrac{q_0 C_{N,\kappa}}{|\partial\Omega|}\ep^{\gamma+N-1}}_{\mathrm{(IV)}} = 0.
\end{equation*}
This always admits one positive solution obtained by balancing terms $\mathrm{(III)}$ and $\mathrm{(IV)}$ and given by
\begin{subequations}
	\begin{equation}
		z \sim z_- := \frac{q_0 C_{N,\kappa}}{6(1+\kappa)|\partial\Omega|}\ep^{\gamma + N - 1}
	\end{equation}
	Another positive solution depends on whether $0\leq\kappa<1$, $\kappa=1$, or $\kappa>1$ and is obtained by balancing terms $\mathrm{(II)}$ and $\mathrm{(III)}$, $\mathrm{(I)}$ and $\mathrm{(III)}$, or $\mathrm{(I)}$ and $\mathrm{(II)}$ respectively. The resulting solution is then given by
	\begin{equation}
		z \sim z_+ := \begin{cases}  \frac{1+\kappa}{1-\kappa},& 0\leq\kappa<1, \\  \sqrt{\frac{2}{q_0}}\ep^{-\gamma/2},& \kappa=1,\\   \frac{\kappa-1}{q_0}\ep^{-\gamma},& \kappa > 1	\end{cases}
	\end{equation}
\end{subequations}
In light of Assumption \ref{ass:shift} we will neglect the solutions corresponding to $z\sim z_+$.

\vspace{2ex}\noindent\textbf{Case II:} Suppose now that $\gamma=0$. The cubic \eqref{eq:cubic-full} then becomes
\begin{equation*}
	\underbrace{6q_0 z^3}_{\mathrm{(I)}} + \underbrace{\bigl( 18q_0+ 6(1-\kappa)  \bigr)z^2}_{\mathrm{(II)}} - \underbrace{6(1+\kappa) z}_{\mathrm{(III)}} + \underbrace{\tfrac{q_0 C_{N,\kappa}}{|\partial\Omega|}\ep^{N-1}}_{\mathrm{(IV)}} = 0.
\end{equation*}
As in Case I above we balance $\mathrm{(III)}$ and $\mathrm{(IV)}$ to get the positive solution
\begin{subequations}
	\begin{equation}
		z \sim z_- := \frac{q_0 C_{N,\kappa}}{6(1+\kappa)|\partial\Omega|}\ep^{ N - 1}.
	\end{equation}
	Moreover, we can find an additional positive solution by balancing terms $\mathrm{(I)}$, $\mathrm{(II)}$, and $\mathrm{(III)}$. This yields a quadratic from which we readily obtain the remaining positive solution
	\begin{equation}\label{eq:z_plus_gam_0}
		z\sim z_+ :=-\left(\frac{1-\kappa}{2q_0} + \frac{3}{2} \right) + \sqrt{\left(\frac{1-\kappa}{2q_0}+\frac{3}{2}\right)^2 + \frac{1+\kappa}{q_0}} 
	\end{equation}
\end{subequations}
In contrast to Case I above, the positive solution $z\sim z_+$ satisfies Assumption \ref{ass:shift} when $\kappa > \kappa_\star$.

\vspace{2ex}\noindent\textbf{Case III:} Finally, we consider the case when $\gamma<0$ for which \eqref{eq:cubic-full} becomes
\begin{equation}
	\underbrace{6q_0z^3}_{\mathrm{(I)}} + \underbrace{18q_0 z^2} _{\mathrm{(II)}}+ \underbrace{\bigl(3\tfrac{q_0C_{N,\kappa}}{|\partial\Omega|}\ep^{N-1}-6(1+\kappa)\ep^{-\gamma}\bigr)z}_{\mathrm{(III)}} + \underbrace{\tfrac{q_0C_{N,\kappa}}{|\partial\Omega|}\ep^{N-1}}_{\mathrm{(IV)}} = 0.
\end{equation}
Notice that $\mathrm{(III)}$ is the only term that may be negative, and furthermore this is possible only when $\gamma \geq  1-N$. We assume for the moment that the inequality is strict and will show that in fact $\gamma \geq \frac{1-N}{2}$ is required in order to  have any positive solutions. In such a case, any positive solution will, to leading order in $\varepsilon\ll 1$, require balancing the negative term $\mathrm{(III)}$. An immediate consequence is that $z\ll 1$. Hence we can neglect term $\mathrm{(I)}$ and this yield a quadratic with roots
\begin{equation}
	z\sim z_{\pm} =  \frac{1+\kappa}{6 q_0}\ep^{-\gamma} \pm \ep^{-\gamma}\sqrt{\biggl(\frac{1+\kappa}{6 q_0}\biggr)^2 - \frac{\ep^{N-1+2\gamma}C_{N,\kappa}}{18|\partial\Omega|}}.
\end{equation}
We immediately see that $\gamma \geq  \tfrac{1-N}{2}$ is necessary to get two positive real roots. Moreover, at the threshold value of $\gamma=\tfrac{1-N}{2}$ we obtain an upper bound for $q_0$ and hence for $q_\ep$. Specifically, we conclude that the cubic \eqref{eq:cubic-full} has exactly two positive solutions provided that
\begin{equation}\label{eq:q-existence-threshold}
	0<q_\ep \leq  q_0^\star \ep^{\tfrac{1-N}{2}},\qquad q_0^\star := (1+\kappa)\sqrt{\tfrac{|\partial\Omega|}{2C_{N,\kappa}}},
\end{equation}
and it has no positive solutions otherwise, which establishes \eqref{eq:princ-result-A-crit-limiting}.

\begin{remark}
	We remind the reader that solutions with $y_{\ep,\bls} \sim -\log z_\pm$ will be referred to as $\mathrm{BLS}_{\pm}$ solutions respectively.
\end{remark}

\subsection{Leading Order Behaviour of the Inhibitor}\label{subsec:leading-order-inhibitor}

We now turn our attention towards determining the leading order behaviour of the inhibitor $\xi_\ep$ given by \eqref{eq:xi_eps_def} . The main idea throughout this calculation is that the contribution of the boundary layer (mediated by $\eta(y_{\ep,\bls})$) relative to that of the interior spike (mediated by $C_{N,\kappa} $) depends on the magnitude of the shift-parameter $y_{\ep,\bls}$. 

Consider first the case of $\BLSm$ solutions when $\gamma>\tfrac{1-N}{2}$. In this case $z\sim z_- = O(\ep^{\gamma+N-1})$ so that \eqref{eq:z-conversion} implies that $\eta(y_{\ep,\bls})=O(\ep^{2N+2\gamma-2})$. Since $\gamma>\tfrac{1-N}{2}$ we deduce that $\ep^{2N+2\gamma-1} \ll \ep^{N}$ and therefore
\begin{equation}
	\xi_\ep\sim \xi_- := \frac{|\Omega|}{C_{N,\kappa}}\ep^{-N}.
\end{equation}
On the other hand, in the case of $\BLSp$ solutions, for any $\kappa\geq 0$ and $\frac{1-N}{2}<\gamma<0$ we find that $\eta(y_{\ep,\bls})\sim 2(\frac{1+\kappa}{q_0})^2 \ep^{-2\gamma}$, and since $\ep^{-2\gamma + 1} \gg \ep^{N}$ we deduce that
\begin{equation}\label{eq:xi_z_plus_gam_neg_range}
	\xi_\ep\sim\xi_+ := \frac{|\Omega|}{2|\partial\Omega|}\left(\frac{q_0}{1+\kappa}\right)^2\ep^{2\gamma - 1}.
\end{equation}
When $\gamma=0$ we must restrict our attention to $\kappa>\kappa_\star$ in order for the $\BLSp$ solution to satisfy Assumption \ref{ass:shift}. In such a case $z\sim z_+ = O(1)$ is given by \eqref{eq:z_plus_gam_0} so that $\eta(y_{\ep,\bls})=O(1)$ and we deduce
\begin{equation}\label{eq:xi_z_plus_gam_0}
	\xi_\ep\sim\xi_+:= \frac{|\Omega|}{|\partial\Omega|}\frac{(z_++1)^3}{6z_+^2(z_++3)}\ep^{-1}.
\end{equation}
In summary, for $\gamma>\tfrac{1-N}{2}$ the dominant contribution to the inhibitor for the $\BLSm$ (resp. $\BLSp$) solution comes from the interior spike (resp. boundary layer). In contrast, when $\gamma=\tfrac{1-N}{2}$ we find that the contribution to the inhibitor from the interior spike and the boundary layer are comparable. Indeed, when $\gamma=\tfrac{1-N}{2}$ we find that $z_\pm = O(\ep^{\tfrac{N-1}{2}})\ll 1$ and hence $\eta(y_{\ep,\bls})\sim 18 z_\pm^2 = O(\ep^{N-1})$ so that
\begin{equation}\label{eq:xi_edge_case}
	\xi_\ep\sim \xi_\pm := \frac{|\Omega|}{18|\partial\Omega|\zeta_\pm^2 + C_{N,\kappa}}\ep^{-N},\quad \zeta_\pm :=  \frac{1+\kappa}{6 q_0}\pm \sqrt{\biggl(\frac{1+\kappa}{6 q_0}\biggr)^2 - \frac{C_{N,\kappa}}{18|\partial\Omega|}}.
\end{equation}

\section{Linear Stability of Boundary-Layer with an Interior Spike}\label{sec:bls_stability}

We next consider the linear stability of the solutions constructed in Section \ref{sec:bls_existence} above. Let $u = u_\ep + e^{\lambda t}\phi(x)$ and $\xi = \xi_\ep + e^{\lambda t}\psi$ so that retaining only linear terms gives
\begin{equation}
	\psi = \frac{2}{(1+\tau\lambda)|\Omega|}\int_\Omega u_\ep\phi dx\sim \frac{2\xi_\ep}{(1+\tau\lambda)|\Omega|}\left( \ep j[\phi] + \ep^N J_{\kappa}[\phi]  \right),
\end{equation}
where we define the linear functionals
\begin{subequations}
	\begin{equation}
		J_{\kappa}[\phi] := \begin{cases} \int_{\mathbb{R}^N_+} W_\kappa(y)\phi(x_0+\ep y)dy, & \kappa\leq\kappa_\star, \\ \int_{\mathbb{R}^N} W(y)\phi(x_0 + \ep y) dy, & \kappa>\kappa_\star.\end{cases}
	\end{equation}
	and 
	\begin{equation}
		j[\phi] := \int_0^\infty w_c(y+y_{\ep,\bls})\int_{\partial\Omega}\phi(\sigma+\ep y \hat{n}_\sigma)d\sigma dy,
	\end{equation}
\end{subequations}
where $\hat{n}_\sigma$ denotes the inward unit normal at $\sigma\in\partial\Omega$. Substituting into \eqref{eq:gm} and keeping only the linear terms then gives
\begin{equation}\label{eq:nlep-full}
	\ep^2 \Delta\phi - \phi +  2(w_0 + \overline{W}_\kappa)\phi - 2\xi_\ep\frac{\ep j[\phi]  + \ep^N J_{\kappa}[\phi]}{(1+\tau\lambda)|\Omega|}(w_0 + \overline{W}_\kappa)^2 = \lambda\phi,\qquad x\in\Omega.
\end{equation}
The relative contributions of the boundary-layer or spike are determined by whether $\kappa\leq\kappa_\star$ or $\kappa>\kappa_\star$ as well as whether the shift parameter is $y_{\ep,\bls}\sim -\log z_+$ or $y_{\ep,\bls}\sim -\log z_-$. In the remainder of this section we catalogue the resulting non-local eigenvalue problems in each of these cases. In all, four distinct cases need to be considered, with the resulting NLEP indicating unconditional linear stability or instability in three of these. Throughout the remainder of this paper we assume that $\tau=0$ so as to avoid oscillatory instabilities and remark that stability should hold more generally provided that $\tau$ is sufficiently small.

\vspace{1ex}
\noindent\textbf{Case A:} Suppose that $y_{\ep,\bls} \sim -\log z_-$, $\gamma>\tfrac{1-N}{2}$, and $\kappa\geq 0$. Then $\xi_\ep \sim \xi_- = \frac{|\Omega|}{C_{N,\kappa}}\ep^{-N}$ and $w_0(x) \sim 6 z_- e^{-\ep^{-1}\dist(x,\partial\Omega)}\ll 1$ throughout $\Omega$. Moreover, since  $z_-=O(\ep^{\gamma+N-1})$ we deduce $j[\phi] = O(\ep^{\gamma+N-1})$ so that the boundary layer contribution in \eqref{eq:nlep-full} is negligible. Introducing appropriate inner variables depending on whether $0\leq\kappa\leq \kappa_\star$ or $\kappa>\kappa_\star$ we obtain the NLEPs
\begin{subequations}
	\begin{equation}
		\begin{cases} \Delta \Phi - \Phi + 2W_\kappa\Phi - 2\frac{\int_{\mathbb{R}^N_+} W_\kappa \Phi dy}{\int_{\mathbb{R}^N_+} W_\kappa^2 dy}W_\kappa^2 = \lambda \Phi, & y\in\mathbb{R}^N_+ \\ -\partial_{y_N} \Phi + \kappa \Phi = 0, & y_N=0, \end{cases}\qquad (0\leq\kappa\leq\kappa_\star),
	\end{equation}
	and
	\begin{equation}
		\begin{cases} \Delta \Phi - \Phi + 2W\Phi - 2\frac{\int_{\mathbb{R}^N} W\Phi dy}{\int_{\mathbb{R}^N} W^2 dy}W^2 = \lambda \Phi, & y\in\mathbb{R}^N \\ \Phi\rightarrow 0, & |y|\rightarrow\infty, \end{cases}\qquad (\kappa>\kappa_\star).
	\end{equation}
\end{subequations}

If $\kappa > \kappa_\star$ then the classical NLEP theory (see for example Theorem 3.1 in \cite{wei_2014_book}) implies that the NLEP admits only eigenvalues with a negative real part. On the other hand, for $\kappa\leq\kappa_\star$ a similar argument (see Appendix \ref{app:half_space_nlep}) likewise implies that the all eigenvalues of the NLEP have negative real part. The solution is therefore linearly stable for all $\kappa\geq 0$.

\vspace{1ex}
\noindent\textbf{Case B:} Suppose now that $y_{\ep,\bls}\sim -\log z_+$, $\gamma =0$, and $\kappa > \kappa_\star$. In this case $z_+ = O(1)$ and  $\xi_\ep=O(\ep^{-1})$ is given by \eqref{eq:xi_z_plus_gam_0}. To leading order \eqref{eq:nlep-full} then becomes
\begin{equation*}
	\ep^2 \Delta\phi - \phi +  2(w_0 + W)\phi - \frac{(z_++1)^3}{3z_+^2(z_++3)|\partial\Omega|}\left( j[\phi]  + \ep^{N-1} J_{\kappa}[\phi] \right)(w_0 + W)^2 = \lambda\phi,\quad x\in\Omega.
\end{equation*}
Seeking an eigenfunction of the form $\phi(x)\sim\Phi(\ep^{-1}(x-x_0))$ we find that $\Phi$ must satisfy
\begin{equation}\label{eq:stability_pde_unstable}
	\Delta\Phi - \Phi + 2 W \Phi = \lambda \Phi,\quad y\in\mathbb{R}^N;\qquad \Phi\rightarrow0,\quad |y|\rightarrow\infty.
\end{equation}
Since this always admits an unstable eigenvalue (see for example Lemma 13.5  in \cite{wei_2014_book}) we deduce that this solution is always linearly unstable.

\vspace{1ex}
\noindent\textbf{Case C:} Next we suppose that $y_{\ep,\bls}\sim-\log z_+$, $\tfrac{1-N}{2}<\gamma<0$, and $\kappa\geq 0$. In this case $z_+ = \frac{1+\kappa}{3 q_0}\ep^{-\gamma}$  and $\xi_\ep = O(\ep^{2\gamma-1})$ is given by \eqref{eq:xi_z_plus_gam_neg_range}. Moreover since $z_+\ll 1$ and hence $w_0=O(\ep^{-\gamma})\ll 1$ in $\Omega$, we deduce  that $j[\phi] = O(\ep^{-\gamma})$. Assuming $\kappa>\kappa_\star$ and seeking a solution of the form $\phi(x)\sim\Phi(\ep^{-1}(x-x_0))$ we recover \eqref{eq:stability_pde_unstable} so that this solution is always unstable. On the other hand, if $0\leq\kappa\leq \kappa_\star$ then seeking an eigenfunction of the form $\phi(x)\sim\Phi(\ep^{-1}(x-x_0))$ gives the NLEP
\begin{equation}
	\Delta\Phi - \Phi  + 2W_\kappa \Phi = \lambda\Phi,\quad y\in\mathbb{R}^N_+;\qquad -\partial_y \Phi +\kappa \Phi = 0,\qquad y_N=0,
\end{equation}
which likewise always has an unstable eigenvalue (see Appendix \ref{app:half_space_nlep} below). Hence the solution in this case is always linearly unstable.

\vspace{1ex}
\noindent\textbf{Case D:} Finally we suppose that $y_{\ep,\bls}\sim-\log z_\pm$, $\kappa\geq 0$, and $\gamma=\tfrac{1-N}{2}$. In this case $z_\pm = \zeta_\pm \ep^{\frac{N-1}{2}}$ where $\zeta_\pm=O(1)$ and $\xi_\ep=O(\ep^{-N})$ are given by \eqref{eq:xi_edge_case}. Since $z_\pm\ll 1$ we have $w_0(x) \sim 6 z_\pm e^{-\ep^{-1}\dist(x,\partial\Omega)} = O(\ep^{\frac{N-1}{2}})$ so that $j[\phi] = O(\ep^{\frac{N-1}{2}})$. The contribution of  $w_0$ and $j[\cdot]$ can then be shown to be negligible for both $0\leq\kappa\leq\kappa_\star$ and $\kappa>\kappa_\star$. Introducing appropriate inner variables in both the $0\leq\kappa\leq\kappa_\star$ and $\kappa>\kappa_\star$ cases then gives the NLEPs
\begin{subequations}
	\begin{equation}
		\begin{cases} \Delta \Phi - \Phi + 2W_\kappa\Phi - 2\chi_\pm \frac{\int_{\mathbb{R}^N_+} W_\kappa \Phi dy}{\int_{\mathbb{R}^N_+} W_\kappa^2 dy}W_\kappa^2 = \lambda \Phi, & y\in\mathbb{R}^N_+ \\ -\partial_{y_N} \Phi + \kappa \Phi = 0, & y_N=0, \end{cases}\qquad (0\leq\kappa\leq\kappa_\star),
	\end{equation}
	and
	\begin{equation}
		\begin{cases} \Delta \Phi - \Phi + 2W\Phi - 2\chi_\pm\frac{\int_{\mathbb{R}^N} W\Phi dy}{\int_{\mathbb{R}^N} W^2 dy}W^2 = \lambda \Phi, & y\in\mathbb{R}^N \\ \Phi\rightarrow 0, & |y|\rightarrow\infty, \end{cases}\qquad (\kappa>\kappa_\star).
	\end{equation}
	where
	\begin{equation}\label{eq:chi_pm_def}
		\chi_\pm := \frac{C_{N,\kappa}}{18|\partial\Omega|\zeta_\pm^2 + C_{N,\kappa}}.
	\end{equation}
\end{subequations}

Both the classical full-space NLEP theory (see Theorem 3.1 in \cite{wei_2014_book}), as well as the half-space NLEP theory discussed in Appendix \ref{app:half_space_nlep} imply that the NLEP is linearly stable provided that $\chi_\pm > 1/2$. Notice that we can rewrite $\chi_\pm$ as
\begin{equation}
	\chi_\pm = \frac{1}{2}\frac{\omega}{1\pm\sqrt{1-\omega}},\qquad \omega:= \frac{C_{N,\kappa}}{18|\partial\Omega|}\left( \frac{6q_0}{1+\kappa} \right)^2.
\end{equation}
Since $q_0\leq q_0^\star$ where $q_0^\star$ is the existence threshold given by \eqref{eq:q-existence-threshold}, we deduce that $0<\omega\leq 1$ and therefore
\begin{equation}
	\begin{cases}
		0\leq \chi_+ \leq \frac{1}{2},\qquad \chi_+|_{\omega=0}=0,\qquad \chi_+|_{\omega=1}=\frac{1}{2}, \\
		\frac{1}{2} \leq \chi_- \leq 1,\qquad \chi_-|_{\omega=0}=1,\qquad \chi_-|_{\omega=1}=\frac{1}{2}.
	\end{cases}
\end{equation}
We thus conclude that the $\BLSp$ solution is always linearly unstable whereas the $\BLSm$ solution is always linearly stable (provided that it exists).

\section{Numerical Simulations}\label{sec:numerical}

We validate the asymptotic analysis of the preceding sections by simulating the time-dependent system \eqref{eq:gm} using the finite element PDE solver FlexPDE 7 \cite{flexpde7}. Throughout our numerical experiments we choose $\Omega\in\mathbb{R}^2$ to be the unit disk, $\ep=0.02$, and $\tau=0$. All asymptotic solutions are computed by directly solving the cubic \eqref{eq:princ-result-cubic} numerically,  including the $\ep$-dependent thresholds $A^\ep_{\text{crit},\bl}$ and $A_{\text{crit},\bls}^\ep$.

\begin{figure}[t!]
	\begin{subfigure}{0.66\textwidth}
		\centering
		\includegraphics[width=\linewidth]{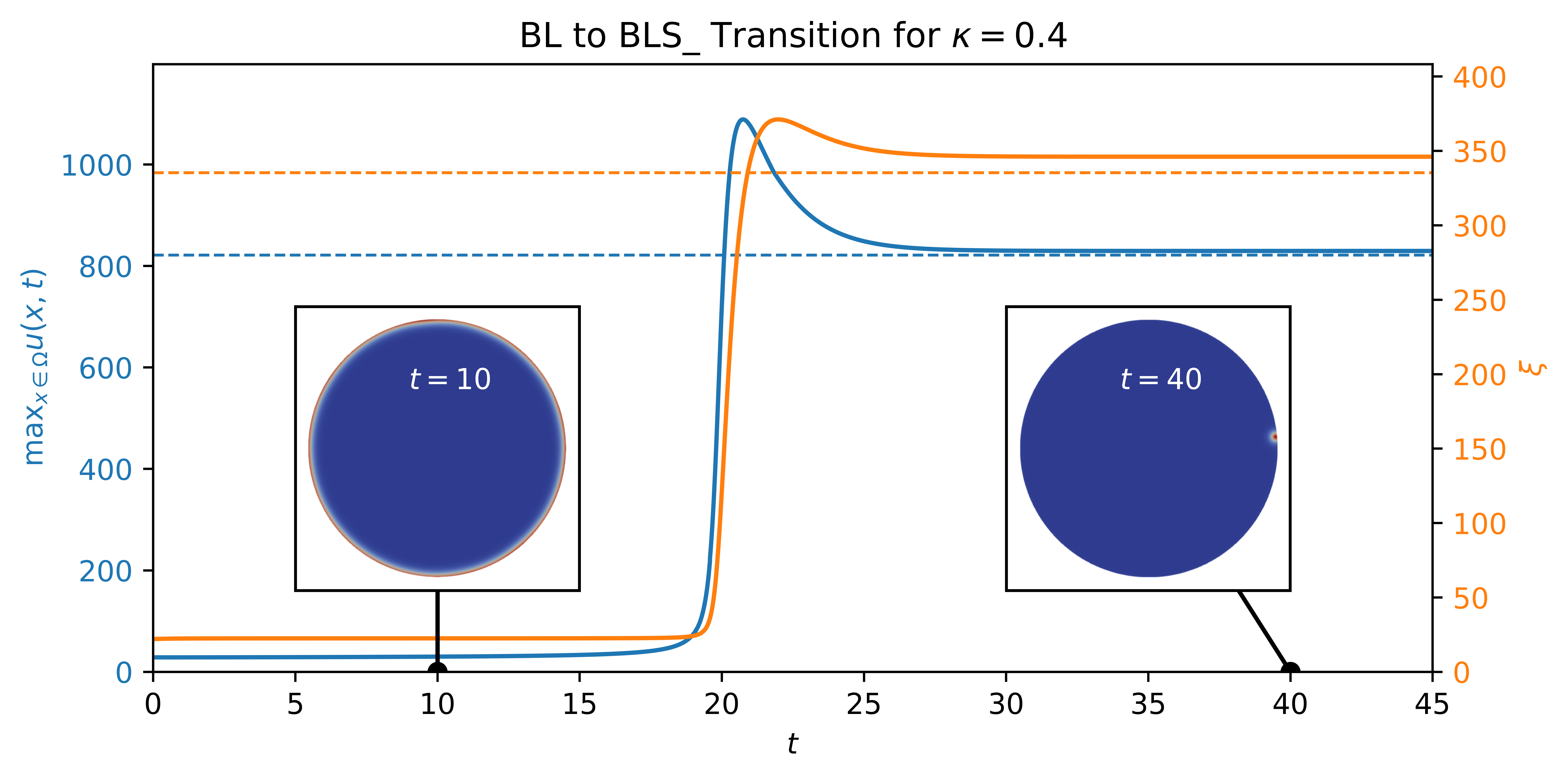}
	\end{subfigure}%
	\begin{subfigure}{0.33\textwidth}
		\centering
		\includegraphics[width=\linewidth]{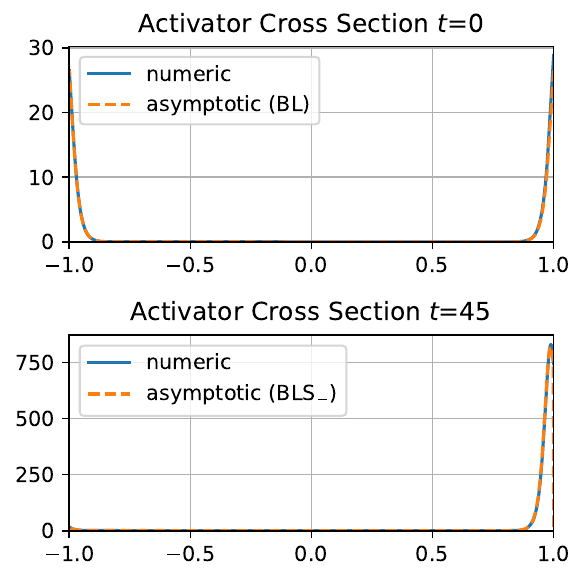}
	\end{subfigure}
	\caption{Numerical simulations illustrating the emergence of a boundary layer with interior spike from the destabilization of a boundary layer when $A=0.95A_{\text{crit,BL}}^\ep$, $\ep=0.02$, and $\tau=0$, and $\kappa=0.4$. In the left plot the blue curve (with corresponding left axis) and orange curve (with corresponding right axis) indicate values of the activator peak value and inhibitor respectively. The dashed blue and orange horizontal lines indicate values predicted by the $\BLSm$ asymptotics. Insets show the activator at $t=10$ and $t=40$. The two right-most plots show cross sections of the activator passing through the spike at $t=0$ (top) and $t=45$ (bottom), comparing numerical results (solid) with the asymptotic solutions (dashed).}\label{fig:bl_bls_kappa_0.4}
\end{figure}

\begin{figure}[t!]
	\begin{subfigure}{0.66\textwidth}
		\centering
		\includegraphics[width=\linewidth]{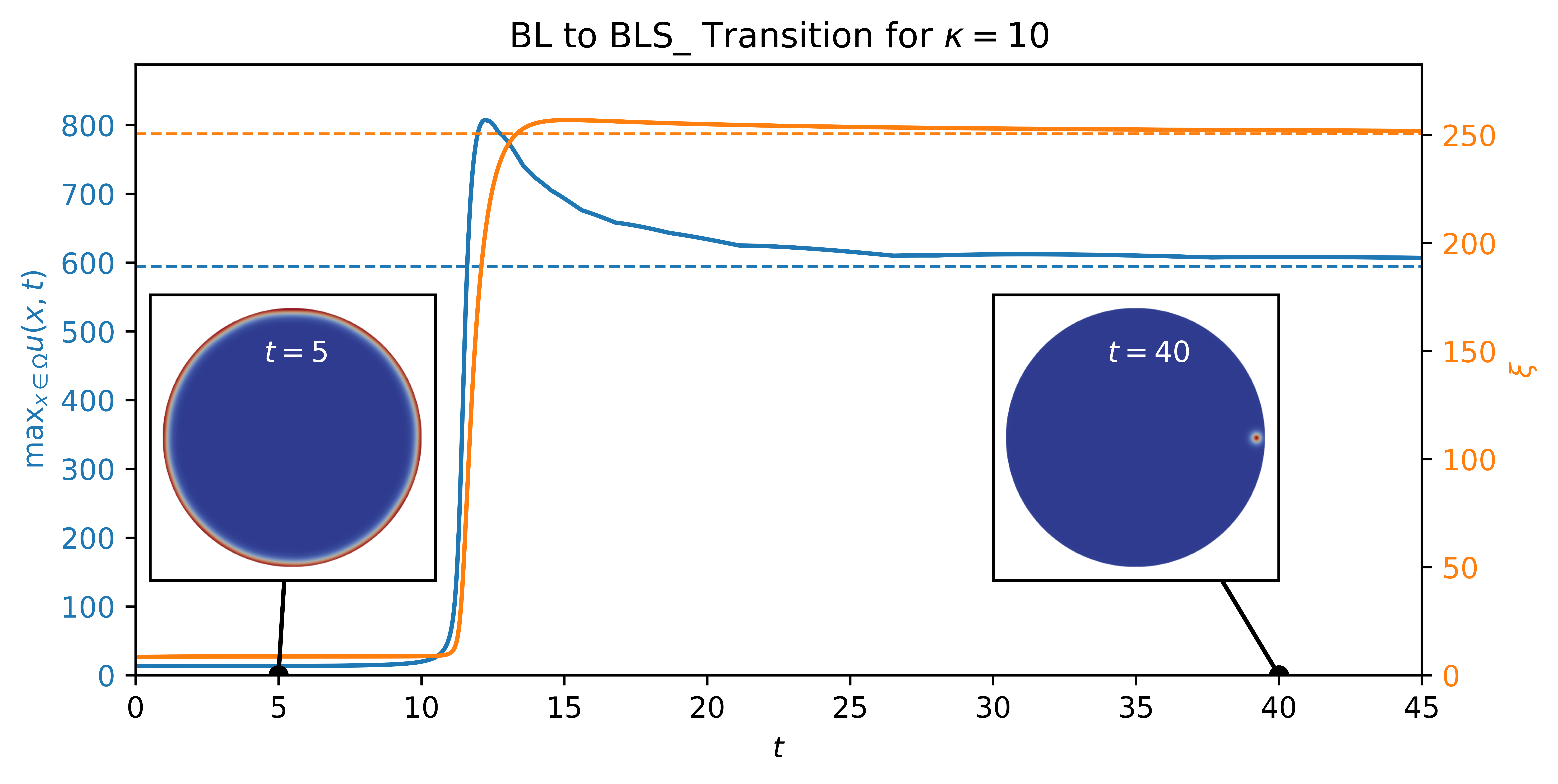}
	\end{subfigure}%
	\begin{subfigure}{0.33\textwidth}
		\centering
		\includegraphics[width=\linewidth]{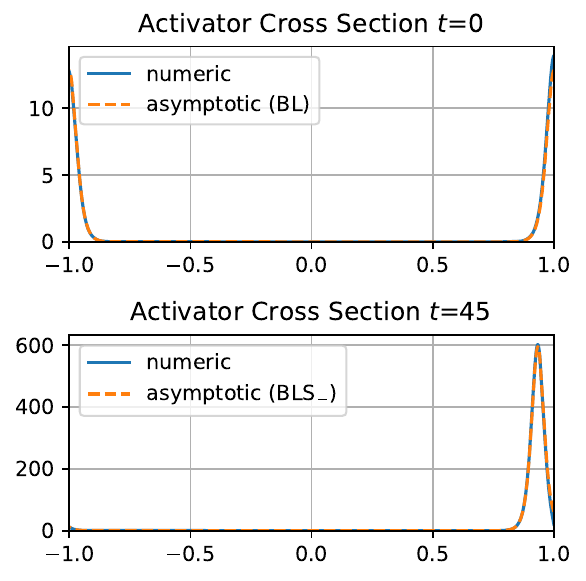}
	\end{subfigure}
	\caption{Description as in Figure \ref{fig:bl_bls_kappa_0.4} with $\kappa=10$.}\label{fig:bl_bls_kappa_10}
\end{figure}

In \cite{gomez_2022} it was previously observed that when $A<A_{\text{crit},\bl}^\ep$ the BL solution is destabilized and transitions to a solution consisting of a boundary layer and an interior spike, which we anticipate corresponds to the $\BLSm$ solution. To support this prediction we perform several simulations starting with the BL solution and a value of $A=0.95A_{\text{crit},\bl}^\ep$. In all cases we find that after the BL solution was destabilized it tends to the $\BLSm$ solution and we illustrate this in Figures \ref{fig:bl_bls_kappa_0.4} and \ref{fig:bl_bls_kappa_10} for $\kappa=0.4$ and $\kappa=10$ respectively. Note that when $\kappa>\kappa_\star$ the spike in the $\BLS$ solution should concentrate at $\text{argmax}_{x\in\Omega}\dist(x,\partial\Omega)$. Our numerical simulations indicate that, upon destabilizing the boundary layer, the interior spike forms near the boundary and then slowly drifts toward the center of the domain.

\begin{figure}[t!]
	\begin{subfigure}{0.66\textwidth}
		\centering
		\includegraphics[width=\linewidth]{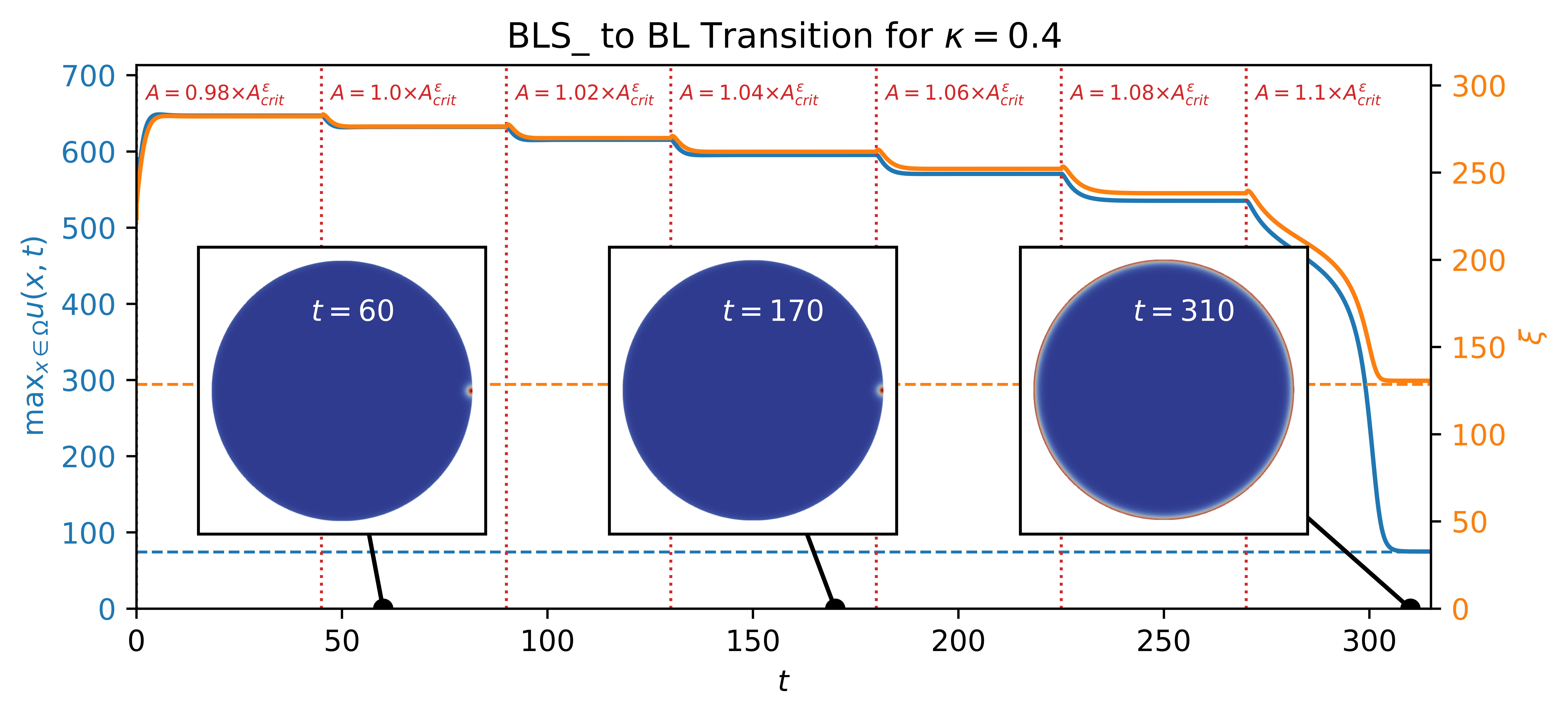}
	\end{subfigure}%
	\begin{subfigure}{0.33\textwidth}
		\centering
		\includegraphics[width=\linewidth]{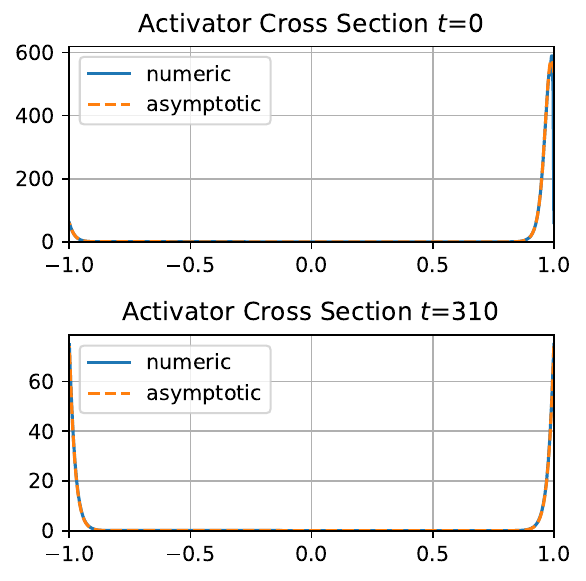}
	\end{subfigure}
	\caption{Numerical simulations illustrating the destabilization of the $\BLSm$ solution as $A$ is increased beyond the existence threshold $A_\text{crit}^\ep$ for $\ep=0.02$, $\tau=0$, and $\kappa=0.4$. When $t=0$ a value of $A=0.98 A_\text{crit}^\varepsilon$ is used and this is increased by $0.02 A_\text{crit}^\varepsilon$ at discrete times indicated by the  vertical red dotted lines in the left plot. In the left plot the blue curve (with corresponding left axis) and orange curve (with corresponding right axis) indicate values of the activator peak value and inhibitor respectively. The dashed blue and orange horizontal lines indicate values predicted by the BL asymptotics.  Insets show the activator at $t=60,170,310$. The two right-most plots show cross sections of the activator passing through the spike at $t=0$ (top) and $t=310$ (bottom), comparing numerical results (solid) with the asymptotic solutions (dashed).}\label{fig:bls_bl_kappa_0.4}
\end{figure}

\begin{figure}[t!]
	\begin{subfigure}{0.66\textwidth}
		\centering
		\includegraphics[width=\linewidth]{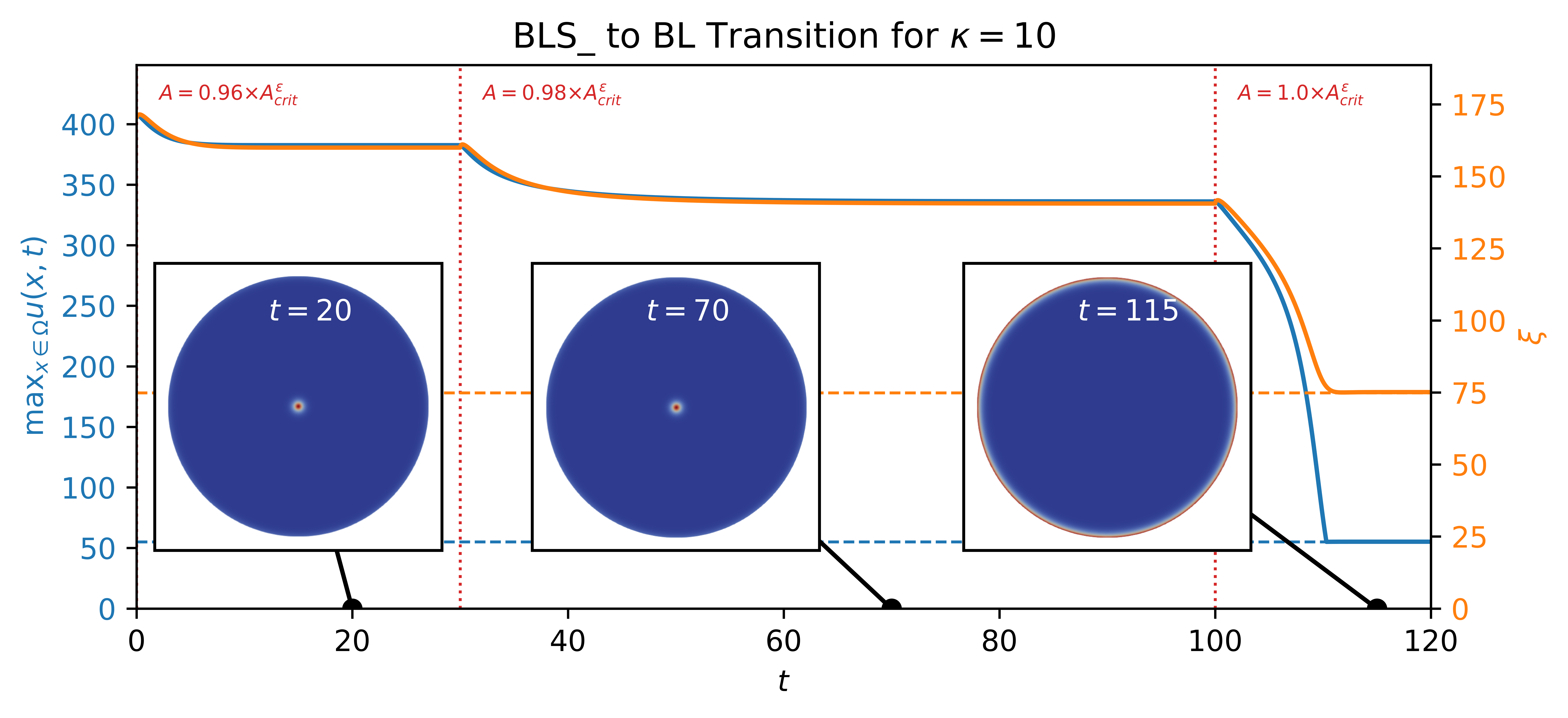}
	\end{subfigure}%
	\begin{subfigure}{0.33\textwidth}
		\centering
		\includegraphics[width=\linewidth]{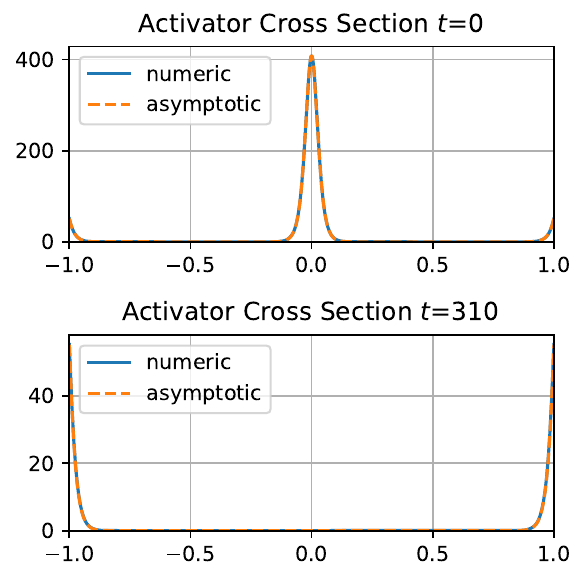}
	\end{subfigure}
	\caption{Description as in Figure \ref{fig:bls_bl_kappa_0.4} with $\kappa=10$.}\label{fig:bls_bl_kappa_10}
\end{figure}

The destabilization of the $\BLSm$ solution coincides with values of $A>A_{\text{crit}}^\ep$ which also corresponds to the existence threshold. Since we don't have a candidate solution beyond this threshold we instead perform numerical simulations in which $A$ is slowly increased beyond the existence threshold. We find that the $\BLSm$ solution is stable when $A<A_{\text{crit}}^\ep$ but transitions to the BL solution when $A$ sufficiently exceeds the threshold $A<A_{\text{crit}}^\ep$. When $\kappa<\kappa_\star$ we find that values of $A\approx 1.1 A_{\text{crit}}^\ep$ are needed to destabilize the $\BLSm$ solution whereas values of $A\approx A_\text{crit}^\ep$ are needed for values of $\kappa>\kappa_\star$. The large error for $\kappa<\kappa_\star$ is likely due to errors in the approximate solution to the interior spike equation \eqref{eq:s-eq}. Specifically, since the spike concentrates near the boundary for $\kappa<\kappa_\star$ there may be a non negligible error from the boundary layer in \eqref{eq:s-eq}. We illustrate the transition from the $\BLSm$ to BL solutions in Figures \ref{fig:bls_bl_kappa_0.4} and \ref{fig:bls_bl_kappa_10} for $\kappa=0.4$ and $\kappa=10$ respectively.

Finally, in all our simulations we observed that the $\BLSp$ solution is linearly unstable. Moreover, we found that in some cases the $\BLSp$ solution collapsed to the BL solution whereas in others it transitioned into the $\BLSm$ solution. A systematic investigation of the dynamics of the $\BLSp$ solution, and in particular whether it leads to a $\BLSm$ or BL solution post-instability, is beyond the scope of this paper.

\section{Conclusion}

In this paper we have used the method of matched asymptotic expansions to construct a solution consisting of a BL and an interior spike to the singularly perturbed shadow GM system in a bounded domain $\Omega\subset\mathbb{R}^N$ ($N\geq 2$). These solutions were previously numerically observed to arise after the destabilization of a BL solution when the flux $A$ is reduced below a certain stability threshold \cite{gomez_2022}. Our results improve on this previous numerical observation by providing an asymptotic characterization of both the structure and linear stability of these emergent solutions. Specifically, in Section \ref{sec:bls_existence} we found that the shadow GM system \eqref{eq:gm} supports two types of solutions consisting of a BL and an interior spike, which we refer to as $\BLSm$ and $\BLSp$ solutions, and which correspond to positive solutions of the cubic equation \eqref{eq:princ-result-cubic}. These solutions exist provided that $A<A_{\text{crit}}^\ep$ where $A_\text{crit}^\ep=O(\ep^{-(N+1)/2})$. In addition, in Section \ref{sec:bls_stability} the linear stability of the $\text{BLS}_{\pm}$ solutions was determined by considering certain full- or half-space NLEPs from which we deduced that the $\BLSp$ solution is always linearly unstable whereas the $\BLSm$ solution is always (provided it exists) linearly stable. Interestingly, the BL solution was previously shown to be linearly stable provided that $A>A_{\text{crit,BL}}^\ep=O(\ep^{-1})$ \cite{gomez_2022} which implies that for $N\geq 2$ there is an asymptotically large range of $A>0$ values over which both the BL solution and the $\BLSm$ solutions exist and are linearly stable.

We conclude with a few suggestions for future research. The first is to extend the present analysis to the case where $\tau$ is larger and for which the $\BLSm$ solution may exhibit a Hopf bifurcation. In this direction it would be interesting to see if oscillatory instabilities can lead to a periodic switching behaviour between the $\BLSm$ and BL solutions that are both linearly stable over the large range $A_\text{crit,BL}^\ep <A<A_\text{crit}^\ep$.  A second collection of open questions involve the dynamics of the $\text{BLS}_\pm$ solutions beyond the onset of instabilities. Specifically, can it be shown that the $\BLSm$ solution jumps to the BL solution as $A$ increases beyond $A_\text{crit}^\ep$? Moreover, it was numerically observed that the $\BLSp$ solution (which is always linearly unstable) sometimes jumps to the $\BLSm$ solution and other times to the BL solution. Is there a threshold value of $A$ below which one behaviour takes place and above which the other? Finally, the present study has considered only the shadow limit for which the inhibitor is well mixed. In the case of homogeneous Neumann or Dirichlet boundary conditions it is known that multi-spike solutions can be sustained for finite values of $D$ \cite{iron_2001}. A natural direction for future work is therefore to consider the case of a finite inhibitor diffusivity and determine the existence and linear stability, paying special attention to the role of the boundary layer, of multi-spike solutions in the case of inhomogeneous boundary conditions considered in this paper.

\section*{Acknowledgments}

D. Gomez was supported by the Simons Foundation Math + X grant and NSERC. J. Wei was partially supported by NSERC.

\addcontentsline{toc}{section}{References}
\bibliographystyle{abbrv}
\bibliography{bibliography}

\appendix

\section{The Half-Space Core Problem}\label{app:half_space_core}

\begin{table}[b!]
	\centering
	\begin{tabular}{|c|c|c|c|c|}
		\hline
		\diagbox{$p$}{$N$} & 2     & 3     & 4     & 5     \\\hline
		2                 & 1.035 & 1.117 & 1.272 & 1.692 \\\hline
		3                 & 1.109 & 1.485 & ----- & ----- \\\hline
	\end{tabular}
	\vspace{1ex}
	\caption{Numerically computed existence threshold $\kappa_\star$ for the half-space core problem \eqref{eq:half-space-core-problem} for select values of $p$ and $N$.}\label{tbl:kappa_star}
\end{table}

Least energy solutions of the half-space core problem \eqref{eq:half-space-core-problem} are expected to give the local profile of equilibrium near-boundary spike solution to \eqref{eq:s-eq} provided that $\kappa$ does not exceed the existence threshold $\kappa_\star>1$ predicted by Theorem 1.1 of \cite{berestycki_2003}. Consider the more general half-space core problem
\begin{equation}\label{eq:half-space-core-problem-general}
	\begin{cases}
		\Delta u - u + u^p = 0,\quad u>0, & \text{in }\, \mathbb{R}^N_+:= \{(y',y_N)\in\mathbb{R}^{N-1}\times\mathbb{R}\,|\, y_N> 0\} \\ u\in H^1(\mathbb{R}^N_+),\quad -\frac{\partial u}{\partial y_N} +\kappa u = 0, &\text{on }\, \partial\mathbb{R}^N_+,
	\end{cases}
\end{equation}
for select values of $p$ and $N$ satisfying $1<p<(N+2)/(N-2)$ if $N\geq 3$ and $p>1$ if $N=2$. The corresponding energy is given by
\begin{equation*}
	I_\kappa[u] = \int_{\mathbb{R}^N_+}\biggl(\frac{1}{2}|\nabla u|^2  + \frac{1}{2}u^2\biggr)dy - \frac{1}{p+1}\int_{\mathbb{R}^N_+}u^{p+1}dy + \frac{\kappa}{2}\int_{\mathbb{R}^N_+} u^2dy.
\end{equation*}
When $\kappa=0$ the least energy solution satisfying \eqref{eq:half-space-core-problem-general} is given by the solution $W$ of the full-space core problem \eqref{eq:full-space-core}. Importantly, denoting by $W_\kappa$ the least energy solution to \eqref{eq:half-space-core-problem-general}, we have the following upper bound(see Section 2 of \cite{berestycki_2003})
\begin{equation}\label{eq:app-upper-bound}
	I_\kappa[W_\kappa]  < 2I_{0}[W],\qquad 0\leq \kappa<\kappa_\star.
\end{equation}
In this appendix we numerically compute solutions to the half-space core problem \eqref{eq:half-space-core-problem-general}. Specifically, we use the $\kappa=0$ solution to initialize a numerical continuation in $\kappa>0$ and use the upper bound \eqref{eq:app-upper-bound} as a stopping criteria with which the critical threshold $\kappa_\star$ can be numerically approximated.

\begin{figure}[t!]
	\begin{subfigure}{0.33\textwidth}
		\centering
		\includegraphics[width=\linewidth]{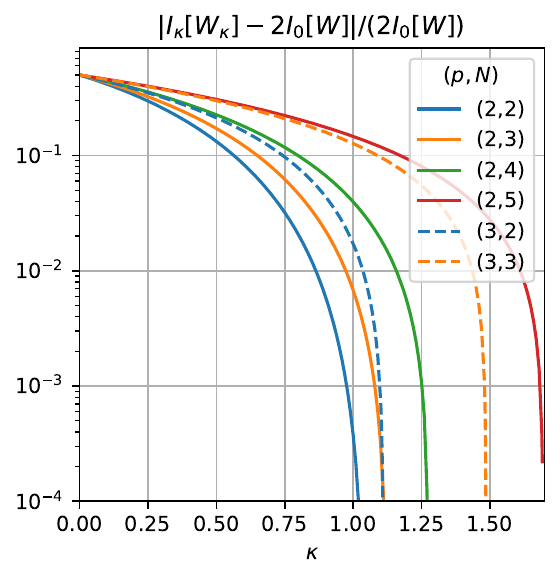}
		\caption{}\label{fig:half-space-core-problem-energies}
	\end{subfigure}%
	\begin{subfigure}{0.33\textwidth}
		\centering
		\includegraphics[width=\linewidth]{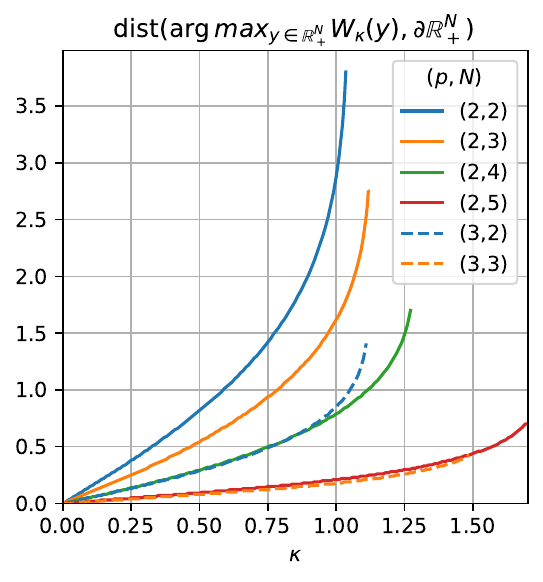}
		\caption{}\label{fig:half-space-core-problem-ymax}
	\end{subfigure}
	\begin{subfigure}{0.33\textwidth}
		\centering
		\includegraphics[width=\linewidth]{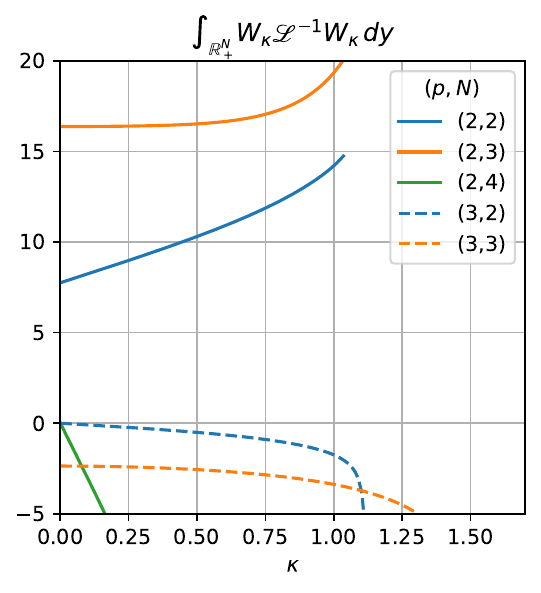}
		\caption{}\label{fig:half-space-core-problem-I11}
	\end{subfigure}
	\caption{(A) Relative difference between energies $I_\kappa[W_\kappa]$ and $2I_0[W]$ as a function of $\kappa$ for select values of $p$ and $N$. (B) Distance from $\partial\mathbb{R}_+^N$ of the half-space core solutions maximum versus $\kappa$ for select values of $p$ and $N$. (C) Plots of $\int_{\mathbb{R}_+^N} W_\kappa\mathscr{L}^{-1} W_\kappa dy$. }
\end{figure}

When $N=1$ the unique radially symmetric least energy solution to \eqref{eq:full-space-core} is explicitly given by
\begin{equation}\label{eq:homoclinic-1D}
	w(y) = \biggl(\frac{p+1}{2}\biggr)^{\tfrac{1}{p-1}}\sech^{\tfrac{2}{p-1}}\biggl(\frac{p-1}{2}y\biggr).
\end{equation}
If instead $N\geq 2$ then this solution must be calculated numerically which, by leveraging its known radial symmetry, reduces to numerically solving the one-dimensional boundary value problem
\begin{equation}\label{eq:full-space-core-radial}
	\begin{cases}
		w'' + (N-1)\rho^{-1}w' - w + w^p = 0,\quad w>0, & \text{in }\rho>0,\\
		w'(0) = 0,\quad w(\rho)\sim C \rho^{-\tfrac{N-1}{2}}e^{-\rho}(1 + O(\rho^{-1})), & \text{as }\rho\rightarrow\infty.
	\end{cases}
\end{equation}
where $\rho=|y|$. We can approximate \eqref{eq:full-space-core-radial} on a truncated domain $0<\rho< L$ with the boundary condition $w'(L)+w(L)=0$. Treating the dimension $N\geq 1$ as a continuous parameter in \eqref{eq:full-space-core} and starting with the known solution \eqref{eq:homoclinic-1D} for $N=1$, we can then slowly increment $N\geq 1$ and use the previously calculated solution as an initial guess with which to solve the next nonlinear boundary value problem. We use this method to calculate the full-space core solutions for given values of $N$ and $p$ by choosing a truncated domain length of $L=20$ and using the \texttt{SciPy} boundary value solver \texttt{solve\_bvp} \cite{2020SciPy-NMeth}.

Next we consider the half-space core problem \eqref{eq:half-space-core-problem-general}. By the moving plane method one can show that solutions to \eqref{eq:half-space-core-problem-general} are in fact symmetric in $y'$ and therefore $u(y) = u(r,y_N)$ where $r=|y'|$. As a consequence we can replace the $N$-dimensional problem \eqref{eq:half-space-core-problem-general} with the two dimensional problem
\begin{equation}\label{eq:half-space-core-problem-general-radial}
	\begin{cases}
		\frac{\partial^2u}{\partial y_N^2} + \frac{1}{r^{N-2}}\frac{\partial}{\partial r}\left(r^{N-2}\frac{\partial u}{\partial r}\right) - u + u^p = 0,\quad u>0, & \text{in }\, r>0,\quad y_N>0, \\ -\frac{\partial u}{\partial y_N} +\kappa u = 0 \quad \text{on }\, y_N=0,\qquad u \rightarrow 0 \quad\text{as }\,r\rightarrow\infty.\\
	\end{cases}
\end{equation}
Letting $L_1>0$ and $L_2>0$ be sufficiently large we seek an approximate numerical solution to \eqref{eq:half-space-core-problem-general-radial} by first introducing the truncated domain $0<r<L_1$ and $0<y_N<L_2$ and then imposing homogeneous Dirichlet boundary conditions on $(r,y_N)\in\{L_1\}\times(0,L_2)$ and $(r,y_N)\in(0,L_1)\times\{L_2\}$ and homogeneous Neumann boundary conditions on $(r,y_N)\in\{0\}\times(0,L_2)$. Letting $\phi$ be a smooth test function vanishing on the boundaries $r=L_1$ and $y_N=L_2$ we obtain the weak formulation
\begin{equation}\label{eq:half-space-core-problem-general-variational}
	\begin{split}
		\int_0^{L_2}\int_0^{L_1}\tilde{\nabla}\phi\cdot\tilde{\nabla}u \;r^{N-2}drdy_N  & + \kappa\int_{0}^{L_1}\phi u \bigr|_{y_N=0} \;r^{N-2} dr \\
		& + \int_0^{L_2}\int_0^{L_1}\phi u \;r^{N-2}drdy_N - \int_0^{L_2}\int_0^{L_1}\phi u^p \;r^{N-2}drdy_N = 0,
	\end{split}
\end{equation}
where $\tilde{\nabla}=\tfrac{\partial^2}{\partial r^2} + \tfrac{\partial^2}{\partial y_N^2}$.

\begin{figure}[t!]
	\centering
	\includegraphics[width=\linewidth]{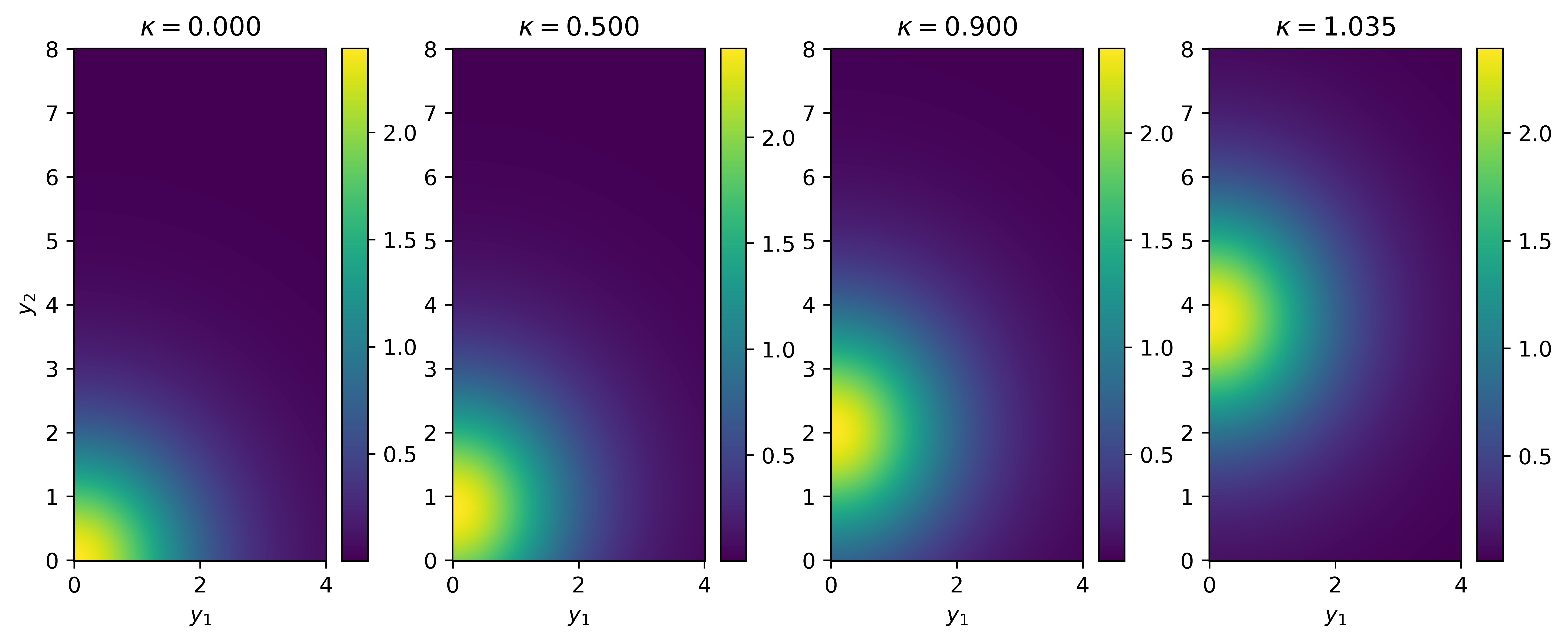}
	\caption{Numerically computed half-space core solution for $(p,N)=(2,2)$.}\label{fig:half-space-core-problem-u_p_2_N_2}
\end{figure}

We solve \eqref{eq:half-space-core-problem-general-variational} numerically by using the finite element method which we implement with FEniCSx \cite{fenicsx-1,fenicsx-2,fenicsx-3}. Specifically, we do this by starting with the numerically calculated solution $W$ of \eqref{eq:full-space-core} when $\kappa=0$ and then slowly incrementing $\kappa\geq 0$, using the previous solution as an initial guess to solve the next nonlinear variational problem \eqref{eq:half-space-core-problem-general-variational}, until (near) equality is reached in \eqref{eq:app-upper-bound}. Using linear Lagrange elements on a structured mesh with $(L_1,L_2)=(10,20)$ consisting of $1000$ and $2000$ nodes in the $x$ and $y$ directions respectively we obtain the numerical approximations to $\kappa_\star$ shown in Table \ref{tbl:kappa_star}. In Figure \ref{fig:half-space-core-problem-energies} we plot $|I_\kappa[W_\kappa] - 2 I_0[W]|/(2 I_0[W])$ as a function of $\kappa$ for select values of $N$ and $p$ which illustrates that near equality in \eqref{eq:app-upper-bound} is reached as $\kappa$ approaches $\kappa_\star$. Additionally, in Figure \ref{fig:half-space-core-problem-ymax} we plot the $y_N$-component of the point where $W_\kappa$ attains its global maximum which shows that this value appears to diverge as $\kappa\rightarrow\kappa_\star$. In Figure \ref{fig:half-space-core-problem-I11} we plot values of $\int_{\mathbb{R}_+^N} W_\kappa\mathscr{L}^{-1} W_\kappa dy$ where the linear operator $\mathscr{L}$ is defined in \eqref{eq:app-L-def} (see Appendix \ref{app:half_space_nlep} for its relevance to the stability of the associated half-space NLEP). Finally, in Figure \ref{fig:half-space-core-problem-u_p_2_N_2} we plot the numerically computed half-space core solution for $(p,N)=(2,2)$ at a sample of $\kappa\leq\kappa_\star$ values.

\section{The Half-Space Non-Local Eigenvalue Problem}\label{app:half_space_nlep}

Let $0\leq \kappa \leq \kappa_\star$. In this appendix we outline the spectral properties of the half-space eigenvalue problem
\begin{equation}\label{eq:app-half-space-eigen}
	\begin{cases}
		\mathscr{L} \Phi = \lambda \Phi, & y\in\mathbb{R}^N_+,\\
		-\partial_{y_N}\Phi +\kappa \Phi = 0, &y_N=0,
	\end{cases}
\end{equation}
and the half-space NLEP 
\begin{equation}\label{eq:app-half-space-nlep}
	\begin{cases}
		\mathscr{L} \Phi - \frac{\mu}{1+\tau\lambda} \frac{\int_{\mathbb{R}^N_+} W_\kappa \Phi dy}{\int_{\mathbb{R}^N_+} W_\kappa^2 dy}W_\kappa^2 = \lambda \Phi, & y\in\mathbb{R}^N_+,\\
		-\partial_{y_N} \Phi + \kappa \Phi = 0, & y_N=0,
	\end{cases}
\end{equation}
where we define the linear operator
\begin{equation}\label{eq:app-L-def}
	\mathscr{L}:= \Delta  - 1 + pW_\kappa^{p-1}.
\end{equation}

We first demonstrate that \eqref{eq:app-half-space-eigen} admits an unstable eigenvalue. Indeed, if $\lambda_0$ is the largest eigenvalue of \eqref{eq:app-half-space-eigen} then 
\begin{equation*}
	\lambda_0 \geq -\frac{\int_{\mathbb{R}^N_+}\{|\nabla W_\kappa|^2+W_\kappa^2-pW_\kappa^{p+1} \}dy + \kappa \int_{\partial\mathbb{R}^N_+}W_\kappa^2dy}{\int_{\mathbb{R}^N_+}W_\kappa^2dy} = (p-1)\frac{\int_{\mathbb{R}^N_+}W_\kappa^{p+1}dy}{\int_{\mathbb{R}^N_+}W_\kappa^2dy} > 0.
\end{equation*}

It is easy to see that $W_\kappa$ satisfies the NLEP \eqref{eq:app-half-space-nlep} with $\lambda=0$ when $\mu=1$. This suggests that $\mu=1$ is, as in the case of the full-space NLEP, the critical threshold for linear stability. In fact, under the following two assumption more can be said.

\begin{assumption}
	The operator $\mathscr{L}$ has a inverse in the class of axially symmetric functions.
\end{assumption}
\begin{assumption}
	The quantity $ \int_{\mathbb{R}_+^N} W_\kappa \mathscr{L}^{-1} W_\kappa dy$ is positive.
\end{assumption}

\begin{theorem*}
	Let Assumptions 1 and 2 above be satisfied.
	\begin{enumerate}
		\item If $\mu<1$ then the NLEP \eqref{eq:app-half-space-nlep} has a positive eigenvalue $\lambda_0>0$.
		\item If $\mu>1$, then there exists a unique $ \tau_c>0$ such that for $\tau <\tau_c$ the NLEP \eqref{eq:app-half-space-nlep} is stable, for $\tau=\tau_c$ it has a pair of purely imaginary eigenvalues, and for $\tau >\tau_c$ it is unstable.
	\end{enumerate}
\end{theorem*}

The proof of this Theorem is similar to that found in \cite{wei_1999}. Additionally, see \cite{maini_2007} for a similar result for the one-dimensional problem with homogeneous Robin boundary conditions, and Sections 3.5 and 3.6 of \cite{wei_2008_handbook} for a discussion of NLEPs with general boundary conditions. We numerically observe that both assumptions required for this theorem hold  for $p=2$ and $2\leq N\leq 3$ (see Figure \ref{fig:half-space-core-problem-I11}) though it remains an open problem to rigorously show this is true.

\end{document}